\documentclass[useAMS,usenatbib]{mn2e}
\usepackage{graphicx}
\DeclareGraphicsExtensions{.ps,.png,.jpg}
\usepackage{txfonts}
\usepackage[T1]{fontenc}
\usepackage{aecompl}
\title[RAFT I: New planetary candidates and updated orbits]
  {RAFT I: Discovery of new planetary candidates and updated orbits from archival FEROS spectra}

\author[M. G. Soto et al.]
  {M. G.~Soto,$^{\star}$
  J. S.~Jenkins,$^{\star}$ M. I.~Jones$^{\dagger}$\\
  $^{\star}$Departamento de Astronom\'ia, Universidad de Chile,
              Camino El Observatorio 1515, Las Condes, Santiago, Chile\\   
              email: msoto\makeatletter @das.uchile.cl\\   
  $^{\dagger}$Department of Electrical Engineering and Center of Astro-Engineering UC, Pontificia Universidad Cat\'olica de Chile,\\
  Av. Vicu\~na Mackenna 4860, 782-0436 Macul, Santiago, Chile}
\date{}

\begin{document}

  \maketitle
   
   \begin{abstract}
A recent reanalysis of archival data has lead several authors to arrive at strikingly different conclusions for a number of planet-hosting candidate stars. In particular, some radial velocities measured using FEROS spectra have been shown to be inaccurate, throwing some doubt on the validity of a number of planet detections. Motivated by these results, we have begun the Reanalysis of Archival FEROS specTra (RAFT) program and here we discuss the first results from this work.  We have reanalyzed FEROS data for the stars HD 11977, HD 47536, HD 70573, HD 110014 and HD 122430, all of which are claimed to have at least one planetary companion. We have reduced the raw data and computed the radial velocity variations of these stars, achieving a long-term precision of $\sim$ 10 m s$^{-1}$ on the known stable star $\tau$ Ceti, and in good agreement with the residuals to our fits.
We confirm the existence of planets around HD 11977, HD 47536 and HD 110014, but with different orbital parameters than those previously published. 
In addition, we found no evidence of the second planet candidate around HD 47536, nor any companions orbiting HD 122430 and HD 70573. 
Finally, we report the discovery of a second planet around HD 110014, with a minimum mass of 3.1 $M_{\mbox{\scriptsize{Jup}}}$ and a orbital period of 130 days.  
Analysis of activity indicators allow us to confirm the reality of our results and also to measure the impact of magnetic activity on our radial velocity measurements.  These results confirm that very metal-poor stars down to [Fe/H]$\sim$-0.7~dex, can indeed form giant planets given the right conditions.
\end{abstract}

\begin{keywords}

techniques: radial velocities -- planets and satellites: general -- planet-star interactions
 
\end{keywords}

%

\section{Introduction}

During the past two decades, the field of exoplanet studies has become one of the dominant areas of astronomy, leading to the discovery of more 1800 planets\footnote{http://exoplanet.eu}. 
Out of the methods used to detect these systems, one of the most successful for detecting planets orbiting the nearest stars is the radial velocity (RV) method, which is performed by measuring the Doppler shift of the spectral lines of a star, produced by the gravitational interaction with orbiting planets. The improvement of instrumentation, calibration techniques, and signal detection methods has made it possible to detect RV signals that conform to planets with masses approaching that of the Earth orbiting the nearest stars (\citealp{alfa_cen_b}; \citealp{jenkins14}; \citealp{anglada14}). 

\citet{bary_corr} has shown that the pipeline of the FEROS spectrograph performs an inaccurate barycentric velocity correction, since it does not include the Earth's precession, introducing a one-year period signal with an amplitude of $\sim$ 62 m s$^{-1}$ for $\tau$ Ceti (a known stable star at the few m s$^{-1}$ level, see \citealt{tuomi13}). In addition, the pipeline uses the initial time of the observation instead of the central 
time, which also introduces another uncertainty, especially for long exposure observations. This issue was also mentioned by \citet{SET2000}.  
This has led some groups to question the detections that were made using FEROS data before this problem was discovered, leading to the reanalysis of data taken with this instrument and the 
rejection of some systems (e.g. HIP 13044 in \citealp{jonesjenkins14}, and HIP 11952 in \citealp{desidera13} and \citealp{bary_corr}).

In this paper we present a reanalysis of FEROS data for the stars HD 11977, HD 47536, HD 70573, HD 110014 and HD 122430, all of which were observed with FEROS and are claimed to have at least one planet orbiting them. Four of these stars are G and K giants and one of them is a very young and nearby star, making them all interesting cases in the study of 
formation and evolution of planetary systems.  

\section{Observations and data reduction}

The data were obtained using the Fiber-fed Extended Range Optical Spectrograph, FEROS (\citealt{kaufer}), mounted on the 2.2m MPG/ESO telescope, at La Silla Observatory. The raw data are available in the ESO archive.\footnote{http://archive.eso.org/eso/eso\_archive\_main.html} The extraction of the FEROS spectra was done with the ESO Data Reduction System (DRS), available for FEROS users. The DRS performs a bias subtraction, flat fielding, order tracing and extraction. The wavelength calibration was computed using two calibration lamps (one ThAr and one ThArNe), having different exposure times, allowing sufficient line coverage across all of the spectral range ($\sim$3500 - 9200 $\AA$). The wavelength solution leads to a typical RMS of 0.005 $\AA$ from $\sim$900 emission lines. Additionally, the reduction pipeline applies a barycentric correction to the extracted data, but this option was disabled and applied later using our own code. 

\subsection{Radial velocity calculation}

The RVs were computed in a similar way as described in \cite{jones13}, according to the following procedure. First, each order was separated into four chunks and the doppler shift was computed by applying a cross correlation (\citealt{crosscorr}) between the stellar spectrum and its corresponding template (high S/N spectrum of the same star). For this purpose we used the IRAF {\it fxcor} task (\citealt{fxcor}). We only used 35 of the 39 orders available, rejecting one in the reddest part of the spectrum (because of the presence of telluric lines) and three in the bluest part of the spectrum (low S/N). 
That left us with a total of 140 chunks, and therefore 140 different velocities per observation. Then, for each RV dataset, the mean velocity was computed, rejecting every point more than 2.5 sigma from the mean. The RVs that were rejected constitute, on average, 17\% of the total number of chunks for each epoch. These rejected chunks were the same for most of the datasets, at the very blue and red ends of the considered spectrum, therefore for most of the epochs we used the same chunks when computing the RVs. The principal cause for rejection of a chunk was because of the low S/N.

Similarly, we computed the cross correlation between the simultaneous calibration lamp (sky fiber) and one of the lamps that was used for the wavelength calibration of that night. The velocity we measured corresponds to the nightly drift that is produced mainly by small pressure and temperature variations inside the spectrograph. 
Finally, we applied the barycentric correction to the measured velocities using the central time of the exposure and the actual coordinates of the star. 

\begin{figure}
\centering
\includegraphics[width=8.8cm]{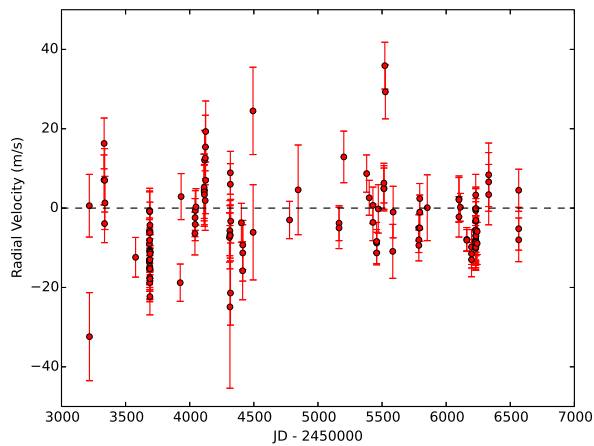}
\caption{ Radial measurements for $\tau$ Ceti from 2004 to 2013. The RMS of the data is 10 m s$^{-1}$.}
\label{HD10700}
\end{figure}

To test our method we calculated the RV of 127 spectra of $\tau$ Ceti, taken from 2004 to 2013. 
The resulting velocities are shown in Figure~\ref{HD10700}. The RMS around zero is $\sim$10 m s$^{-1}$, which is higher than the $\sim$ 4 m s$^{-1}$ obtained by
\citet{jones13} for the same star, however the difference is due to the limited data used in their work that spanned just over 3 years and were observed after 2007.  
Prior to this date there was no standard calibration plan for FEROS, thus the wavelength solution was poorer. 

\subsection{Data from other instruments}

In this work we also supplemented our data with spectra taken with other instruments, namely HARPS, CORALIE and CHIRON. The High Accuracy Radial velocity Planets Searcher, HARPS (\citealt{MAYOR2003}), is located at the 3.6m ESO telescope at La Silla Observatory. Using this spectrograph, \citet{bary_corr} obtained a RMS of 6.9 m s$^{-1}$ for $\tau$ Ceti, proving the accuracy of its measurements. The data was obtained from the ESO Reduced Spectra.\footnote{http://archive.eso.org/wdb/wdb/adp/phase3\_spectral/form?

phase3\_collection=HARPS}
For each star with HARPS data, the mean value of the RVs was subtracted for each epoch, obtaining velocities distributed around zero m s$^{-1}$. We did this to later combine the HARPS RVs with data from other instruments, which were computed using a different template.

CORALIE (\citealt{queloz00}) is at the 1.2m Swiss telescope located at the La Silla Observatory. In \citet{segransan10} they show that 40\% of the measurements made with this instrument reach a radial velocity accuracy of $\sim$ 6 m s$^{-1}$ or better, and 90\% with an accuracy better than $\sim$ 10 m s$^{-1}$. We obtained the CORALIE RV's from the Systemic Console, which were mentioned in \citet[][hereafter S08]{setiawan_HD47536_2} but were never published.

Finally, we used data from the CHIRON spectrograph (\citealt{TOKOVININ2013}), located at the 1.5m telescope at Cerro Tololo Inter-American Observatory (CTIO). The extraction of the data was done using the automatic reduction pipeline that is offered to CHIRON users. The velocities were obtained using the I$_2$ cell technique (\citealt{butler96}). This consists of passing the stellar light through a cell that contains iodine vapor, that superimposes absorption lines in the region between 5000 $\AA$ and 6000 $\AA$. These lines are then used as markers against which the doppler shift can be measured. More detail can be found in \citet{jones14}. With this instrument following our procedures it is possible to reach a velocity precision of $\sim$ 6 m s$^{-1}$.

The fits for the RV curves were made using the Systemic Console (\citealt{console}). We changed the objective function of the minimizer to be the RMS of the fit, instead of the $\chi^2$. 
This was done because of the difference between the uncertainties obtained in the measurements for each instrument. 
The HARPS data have uncertainties on the order of $\sim$ 0.5 m s$^{-1}$, whereas for the other datasets they are generally higher than 5 m s$^{-1}$. 
When we tried to minimize the $\chi^2$, the HARPS data would dominate the fitting process, even when most of the RVs were computed from FEROS spectra.
Minimizing the RMS prevents this from happening, making the fit more evenly distributed across all the datapoints.
The uncertainties in the derived orbital parameters for each star were obtained using the boot strapping option available in the Systemic Console.

\section{Radial velocity analysis}

\subsection{HD 11977}

In \citet[][hereafter S05]{setiawan_HD11977}, this star was classified as a G5III star, located in the ''clump region''. They derived a mass of 1.91 $M_{\odot}$, 
a rotational period of $P_{\mbox{rot}}/\sin i =$ 230-270 days, and a metallicity of [Fe/H] = -0.21. Assuming negligible mass loss during the red giant branch phase, 
it is likely a former main-sequence A-star. We will adopt a mass of 2.31 $M_{\odot}$ and metallicity of [Fe/H] = -0.16, which were derived in \citet[][hereafter M13]{mortier2013}.
S05 also announced a planetary companion with an orbital period of P = 711 days, an eccentricity $e$ = 0.4, and minimum mass 
$m_2\sin i$\,=6.5 $M_{\mbox{\scriptsize{Jup}}}$, being the first planet found orbiting an intermediate mass star (for $M_{\star}=1.19 M_{\odot}$). 

\begin{figure}
\centering
\includegraphics[width=8.8cm]{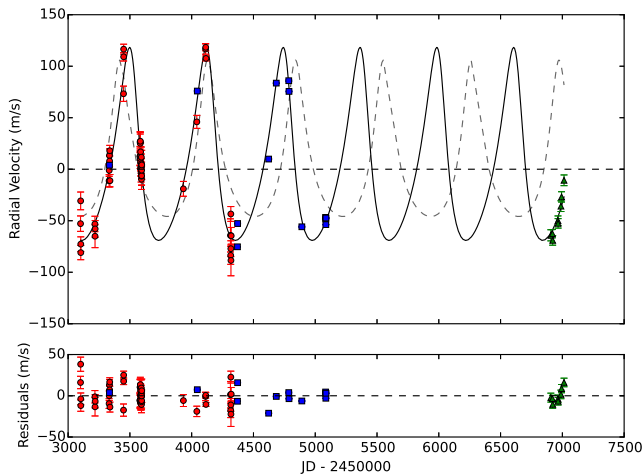}
\caption{ {\it Top panel}: Radial measurements for HD 11977 taken with FEROS (red circles), HARPS (blue squares), and CHIRON (green triangles). The back solid line corresponds to the RV curve found in this paper, with a RMS of 11.2 m s$^{-1}$. The dashed grey line corresponds to the fit found in S05, which gives us a RMS of 33.8 m s$^{-1}$. {\it Bottom planel}: Residuals from our fit.}
\label{HD11977_main}
\end{figure}

\begin{figure}
\centering
\includegraphics[width=8.8cm, height=6cm]{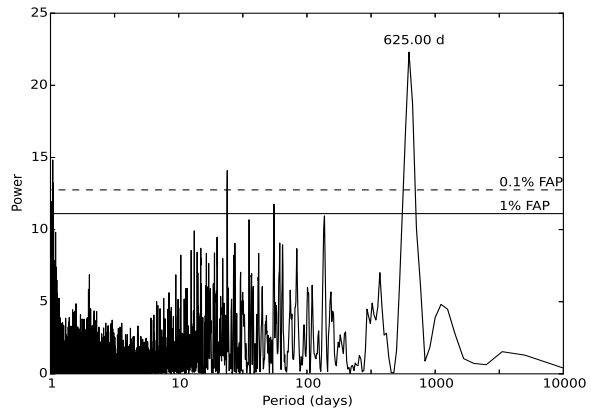}
\caption{LS periodogram of our RV data for HD 11977. The two horizontal lines represent the 1\% (solid line) and 0.1\% (dashed line) FAP.}
\label{HD11977_periodogram}
\end{figure}

\begin{table}
\caption{Orbital parameters for HD 11977 b}           
\label{HD11977b}      
\centering                          
\begin{tabular}{l l}        
\hline\hline                
$P$ (day) &621.6 $\pm$ 2.4\\    
$T_0$ (JD-2450000) & 2906.6 $\pm$ 10\\
$e$ & 0.3 $\pm$ 0.03\\
$\omega$ (deg)& 35.5 $\pm$ 5.8\\
$M_p\sin i$ ($M_{\mbox{\scriptsize{Jup}}}$) & 6.5 $\pm$ 0.2\\
$a$ (AU)& 1.89 $\pm$ 0.005\\
\hline                                  
\end{tabular}
\end{table}

To test this Keplerian solution we used FEROS data of this star taken between
2001 and 2007. However, when testing the nights that were observed before 2004 we obtained a large scatter in the velocities for $\tau$ Ceti ($\sim$ 30 m s$^{-1}$), making those nights 
unreliable. Therefore, we only used the nights from 2004 to 2007, which includes a total of 48 spectra. We also included 217 measurements taken with HARPS.
Since many of these RV epochs were obtained during a single night, we decided to use the RV average in that night, leaving us with 13 RV epochs.
We also included 8 observations made with CHIRON. 
Figure~\ref{HD11977_main} shows the resulting RVs from FEROS (red circles), HARPS data (blue squares), and CHIRON (green triangles). 
When fitting the orbital solution found in S05, we got a RMS of 33.8 m s$^{-1}$, larger than the one published in S05 of 29.1 m s$^{-1}$. 
Figure~\ref{HD11977_periodogram} shows the Lomb-Scargle periodogram (\citealt{periodogram}) of the data. As can be seen, there is a strong peak at 625 days. After close inspection, it was found that this period does not correspond to any peak in the periodogram of the sampling, however we did find that the second peak at 23 days is a product of the sampling of the data.
We fitted a single-planet Keplerian solution to our data, starting from the 625 day period found in the periodogram.
The best fit leads to the following parameters: P = 621 days, $M_p \sin i$ = 6.5 $M_{\mbox{\scriptsize{Jup}}}$, and an eccentricity of $e$ = 0.3. 
The orbital parameters are listed in Table~\ref{HD11977b}. 
This solution leads to a RMS of 11.2 m s$^{-1}$, almost a factor three better than the 29.1 m s$^{-1}$ uncertainty found in S05.

\subsection{HD 47536}\label{hd47536}

The second star in our sample is HD 47536, a very metal-poor K1-III star. \citet{setiawan_HD47536_1} derived a mass of 1.1-3.0 $M_{\odot}$, a rotational 
period of $P_{\mbox{rot}}/\sin i =$ 619 days, and a metallicity of [Fe/H]= -0.68 dex. It was first studied using data from 1999 to 2002 taken with FEROS, and one night with CORALIE in October 2002. 
They reported the discovery of one companion with a period of 712 days and a minimum mass of 5.0  $M_{\mbox{\scriptsize{Jup}}}$. Later, S08 did a revision of the orbital solution including new data from 2004.
They claimed that the star actually hosts two planets with the following orbital parameters: P $\sim$430 days and M$_p$ $\sim$5 $M_{\mbox{\scriptsize{Jup}}}$ for 
the inner planet, and a P $\sim$2500 days and a M$_p$ $\sim$7 $M_{\mbox{\scriptsize{Jup}}}$ for the outer object. 
It is worth mentioning that they did not publish the exact results from the two-planet solution, the values we quote came from an unrefereed conference proceeding.

\begin{figure}
\centering
\includegraphics[width=8.8cm]{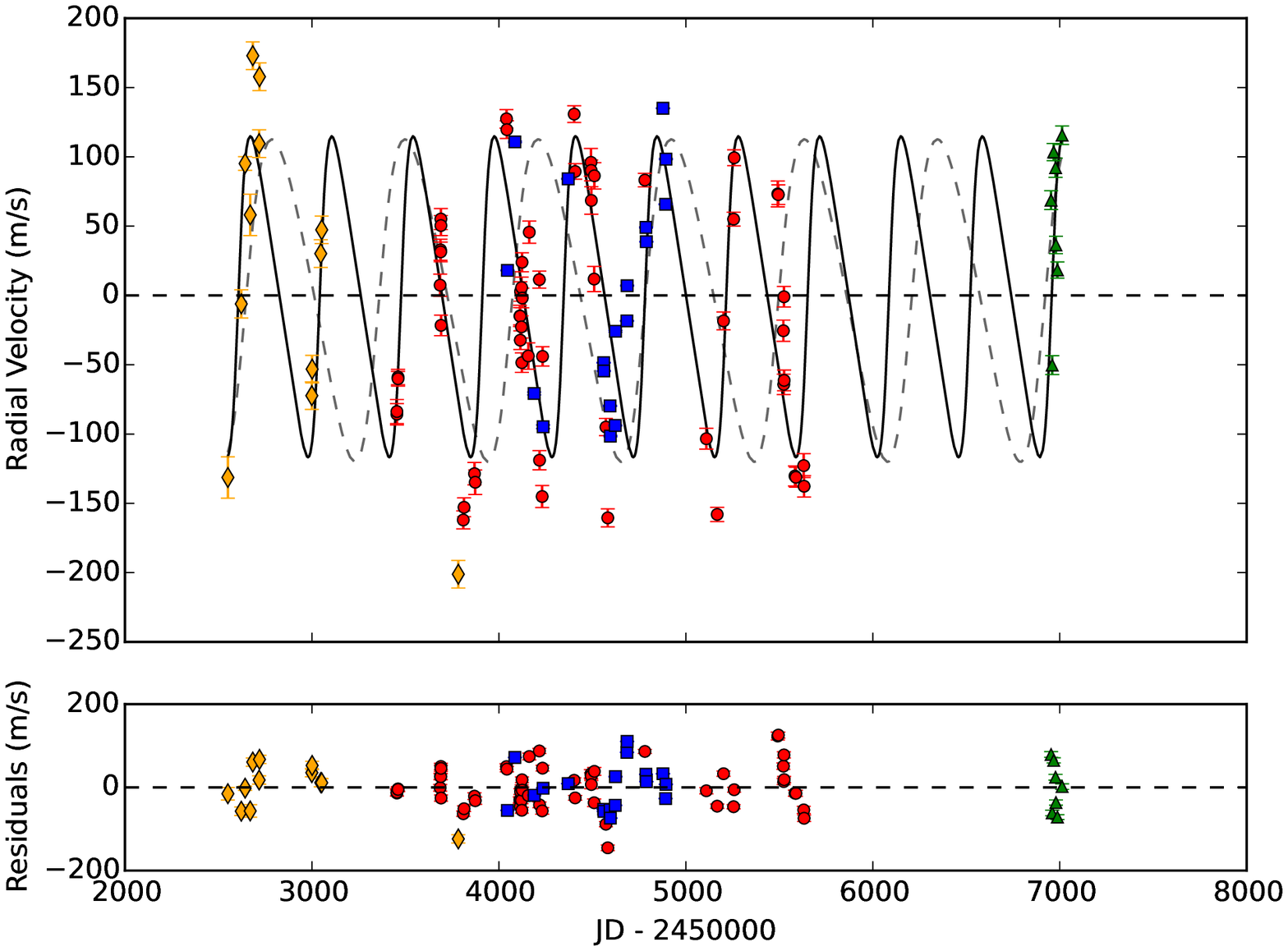}
\caption{{\it Top panel}: Radial velocity measurements for HD 47536 from FEROS (red circles), CORALIE (orange diamonds), HARPS (blue squares), and CHIRON (green triangles). The back solid line corresponds to the RV curve found in this paper. The RMS of this fit is 51.7 m$\mbox{s}^{-1}$. The dashed grey line corresponds to the fit found in \citet{setiawan_HD47536_1}, with a RMS of 103.5 m s$^{-1}$. {\it Bottom panel}: Residuals from the fit found in this paper.}
\label{HD47536_1p}
\end{figure}

\begin{figure}
\centering
\includegraphics[width=8.8cm, height=6cm]{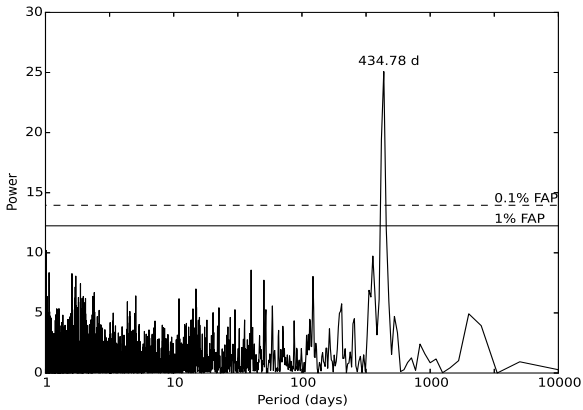}
\caption{Lomb-Scargle periodogram of the HD 47536 RV measurements. The two horizontal lines represent the 1\% (solid line) and 0.1\% (dashed line) FAP.}
\label{HD47536_periodogram}
\end{figure}

\begin{table}
\caption{Orbital parameters for HD 47536 b}            
\label{HD47536b}      
\centering                         
\begin{tabular}{l l}        
\hline\hline                
$P$ (day) &434.9 $\pm$ 2.6 \\    
$T_0$ (JD-2450000)& 3040.8 $\pm$ 30\\
$e$ & 0.3 $\pm$ 0.1 \\
$\omega$ (deg)& 268.6 $\pm$ 17.3 \\
$M_p\sin i$ ($M_{\mbox{\scriptsize{Jup}}}$)& 4.0 $\pm$ 0.4\\
$a$ (AU)& 1.12  $\pm$ 0.005\\
\hline                                   
\end{tabular}
\end{table}
 
We analyzed 56 FEROS spectra of HD 47536, taken from 2005 to 2011. 
We did not include the data taken before 2005 because of the large RV scatter we obtained for $\tau$ Ceti, as was the case for the nights before 2004 for HD 11977. 
Additionally, we included 12 RV epochs obtained with CORALIE, 6 data points obtained with CHIRON, and 18 RVs computed from HARPS spectra. 
The resulting RV measurements are shown in Figure~\ref{HD47536_1p}. The red circles, orange diamonds, blue squares, and green triangles correspond to 
FEROS, CORALIE, HARPS, and CHIRON data, respectively.
We also used the mass and metallicity derived by M13 for this star, corresponding to $M_{\star}=0.98 M_{\odot}$ and [Fe/H]=-0.65 dex.
The LS periodogram of the RVs (Figure~\ref{HD47536_periodogram}) shows a strong peak at 434 days. 
Starting from this period we fit a Keplerian solution to the data, which leads us to the following orbital parameters: P=434.9 days, $e$ = 0.3 and M$_P$ = 4.0 
$M_{\mbox{\scriptsize{Jup}}}$. The orbital solution is overplotted in Figure~\ref{HD47536_1p} (solid line) and we obtain a RMS around the fit of
51.7 m$\mbox{s}^{-1}$. The orbital parameters are listed in Table~\ref{HD47536b}.

\begin{figure}
\centering
\includegraphics[width=8.8cm, height=6cm]{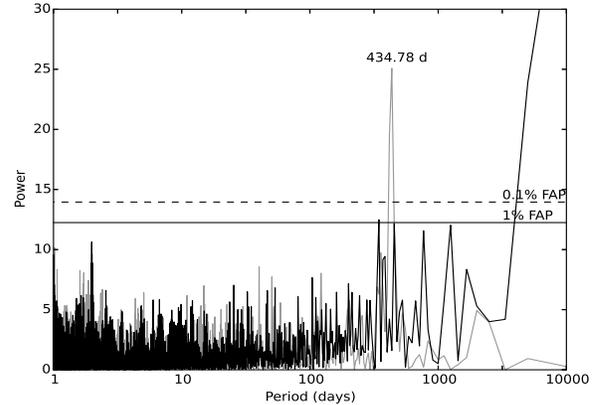}
\caption{Lomb-Scargle periodogram of the sampling window (black line) and the data (grey line) for HD 47536. The peak at 434.78 days corresponds to the peak found in Figure~\ref{HD47536_periodogram}.}
\label{HD47536_sampling}
\end{figure}

In Figure~\ref{HD47536_sampling} the periodogram for window function and the data are shown. We see that there is a peak in the sampling, at 454 days, very close to the peak found in Figure~\ref{HD47536_periodogram}. In order to discard the possibility that the 434-days period is not a product of the sampling of the data, we measured its power in the periodogram as we add data points. If the period is from a true signal from the star, then we expect that it should become stronger as we add more RV measurements, in a monotonic fashion. If it is a product of the sampling, its power could stop growing and will decline in regions. We found that the power grows as we add more data points more or less monotonically, instead of decreasing, which helps to confirm that the 434-days period is produced by a true signal from the system.

\begin{figure}
\centering
\includegraphics[width=8.8cm]{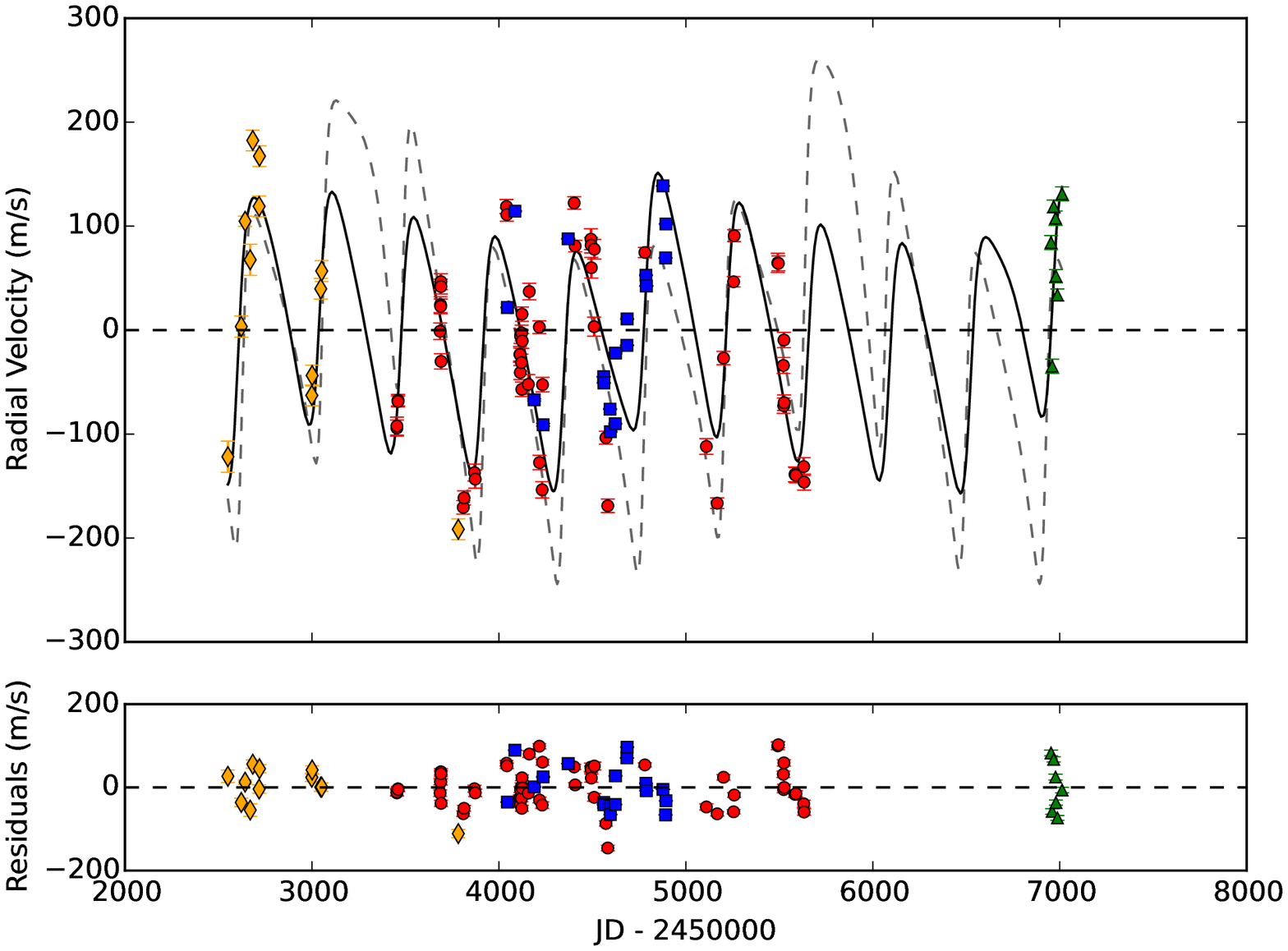}
\caption{{\it Top panel}: Radial velocity for HD 47536 from FEROS (red circles), CORALIE (orange diamonds), HARPS (blue squares), and CHIRON (green triangles). The black solid line corresponds to the RV curve found in this paper for a 2-planet fit, with a RMS of 48.9 m s$^{-1}$. The dashed grey line corresponds to the fit found in S08, with a RMS of 58.8 m s$^{-1}$. {\it Bottom planel}: Residuals from the fit found in this paper.}
\label{HD47536_2p}
\end{figure}

\begin{figure}
\centering
\includegraphics[width=8.8cm, height=6cm]{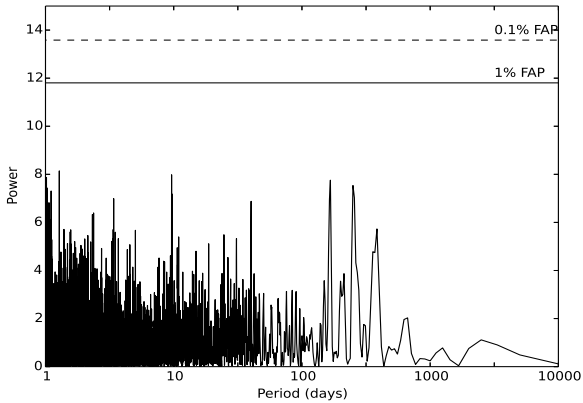}
\caption{Periodogram of the residuals of HD 47536 after fitting the one-planet solution. The two horizontal lines represent the 1\% (solid line) and 0.1\% (dashed line) FAP.} 
\label{HD47536_periodogram_residuals}
\end{figure}

We then tried to fit a two-planet solution using as initial guesses the values for the period and mass of the planets found in S08 (limiting the eccentricity of the outer 
planet to be smaller than 0.5 otherwise the system might be highly unstable). We found a mass for the outer planet of $\sim$ 2.0 $M_{\mbox{\scriptsize{Jup}}}$, with a period of 
2013 days and eccentricity of 0.5. This two-planet fit is shown in Figure~\ref{HD47536_2p}. The RMS of the solution is 48.9 m s$^{-1}$, which is slightly better 
than the one-planet fit. We performed an F-test between these two solutions and found that the two-planet solution is not statistically superior to the one-planet fit 
(probability of 70 \%).

When looking again at the periodogram of the data, we were not able to find any other significant peak that could justify another signal. 
The periodogram of the residuals from the one-planet fit (Figure~\ref{HD47536_periodogram_residuals}) does not show a significant peak neither. 
Therefore, we can rule out the existence of a second statistically significant signal in the current RV data of this star.

\subsection{HD 110014}\label{hd110014}

HD 110014 is a K2III star with an absolute magnitude $M_v$ = -0.11, color ($B-V$) = 1.24, and metallicity [Fe/H]=0.14 dex (M13). In \citet[][hereafter M09]{HD110014} they estimated the upper limit of the rotational period as $P_{\mbox{rot}}/\sin i \sim$ 513 days. They also estimated the mass and age of this star by its position in the color-magnitude diagram, using the process explained in \cite{daSilva06}. According to their solution, HD 110014 could be either a 1.9 $M_{\odot}$ star at the end of its core helium phase, or a 2.4 $M_{\odot}$ star in its first-ascent of the RGB, close to the core helium flash. In this paper we will adopt the mass of 2.09 $M_{\odot}$, as derived by M13. In M09 this star was observed using FEROS from 1999 to 2007, plus a few nights with HARPS and CORALIE. They claim the detection of a planet with a period of 835.5 days, a minimum mass of 11.1-9.5 $M_{\mbox{\scriptsize{Jup}}}$ (for $M_{\star}=1.9-2.4$ $M_{\odot}$), and an eccentricity of 0.46. 

We analyzed spectra taken with FEROS from 2004 to 2011, corresponding to 21 nights and a total of 25 data points. We also used 116 data points taken with HARPS, but as with HD 11977, many of them were taken in the same night and so we took the average velocity per night. That left us with 17 HARPS data points.

\begin{figure}
\centering
\includegraphics[width=8.8cm, height=6cm]{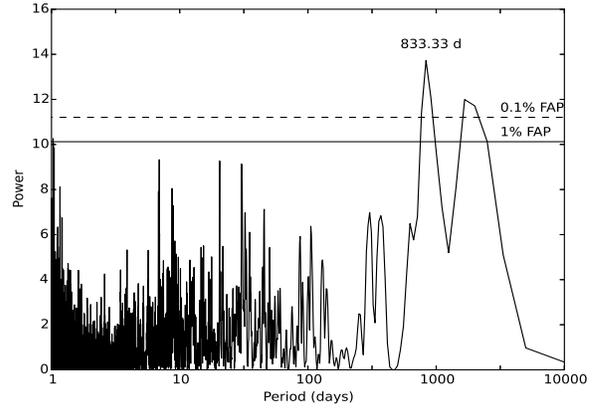}
\caption{LS periodogram of HD 110014. The two horizontal lines represent the 1\% (solid line) and 0.1\% (dashed line) FAP.}
\label{HD110014_periodogram}
\end{figure}

\begin{figure}
\centering
\includegraphics[width=8.8cm]{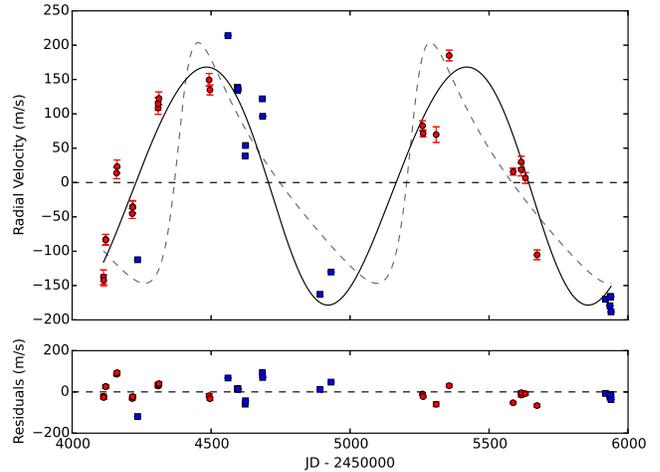}
\caption{{\it Top panel}: Radial measurements for HD 110014 from FEROS (red circles) and HARPS (blue squares). The back solid line corresponds to the RV curve found in this paper, with a RMS of 44.6 m s$^{-1}$. The dashed grey line correspond to the fit found in M09, which gives us a RMS of 63.13 m s$^{-1}$. {\it Bottom planel}: Residuals from the fit.}
\label{HD110014_1p}
\end{figure}

\begin{figure}
\centering
\includegraphics[width=8.8cm, height=6cm]{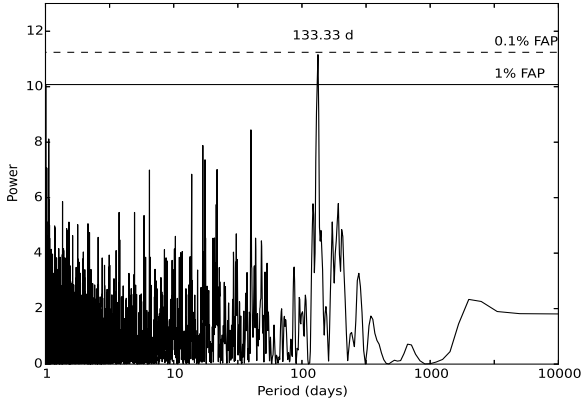}
\caption{Lomb-Scargle periodogram of the residuals of HD 110014, after fitting the one-planet solution. The two horizontal lines represent the 1\% (solid line) and 0.1\% (dashed line) FAP.} 
\label{HD110014_periodogram_residuals}
\end{figure}

\begin{figure}
\centering
\includegraphics[width=8.8cm]{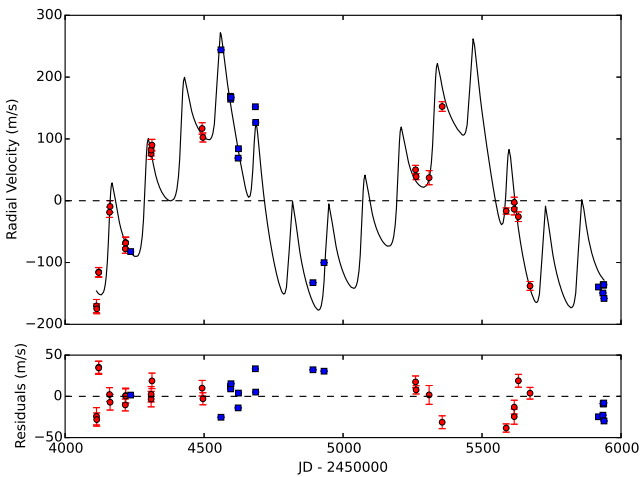}
\caption{{\it Top panel}: Radial velocity measurements for HD 110014 from FEROS (red circles) and HARPS (blue squares). The back solid line corresponds to the RV curve found for a two-planets system. The RMS of this fit is 19.4 m s$^{-1}.$ {\it Bottom planel}: Residuals from the fit.}
\label{HD110014_2p}
\end{figure}

We tried to fit the solution found in M09 to our data, obtaining a RMS of 63.13 m s$^{-1}$, larger than the one published of only 45.8 m s$^{-1}$.  When trying to find another orbital solution we looked at the periodogram of the data in Figure~\ref{HD110014_periodogram}, which shows an emerging peak at 833 days. Beginning with that initial period and minimizing all the orbital parameters, we obtained a fit with a RMS of 44.64 m s$^{-1}$. The fit is shown in Figure~\ref{HD110014_1p} and the orbital parameters are listed in Table~\ref{HD110014_table1}. The period of this possible companion is 936 days, slightly larger than the 833-days period found in Figure~\ref{HD110014_periodogram}. The periodogram of the residuals (Figure~\ref{HD110014_periodogram_residuals}) shows a possible peak at 133 days and is comfortably above the 0.1\% FAP. When a new signal with a period of 133 days is added to our previous one-planet solution, we find a new model as shown in Figure~\ref{HD110014_2p}, and this leads to the orbital parameters listed in Table~\ref{HD110014_table2}. The RMS of this two-planet solution is 19.4 m s$^{-1}$, significantly lower than the one found in M09 of 45.8 m s$^{-1}$ and our own one-planet solution. 

\begin{table}
\caption{Orbital parameters for HD 110014 b.}            
\label{HD110014_table1}      
\centering                          
\begin{tabular}{l l}        
\hline\hline                 
$P$ (day) & 936.4 $\pm$ 44.5\\   
$T_0$ (JD-2450000)& 3842.0 $\pm$ 206.6\\
$e$ & 0.06 $\pm$ 0.09 \\
$\omega$ (deg)& 119.55 $\pm$ 84.2 \\
$M_p\sin i$ ($M_{\mbox{\scriptsize{Jup}}}$)& 13.67 $\pm$ 2.3 \\
$a$ (AU)& 2.5 $\pm$ 0.08\\
\hline                                  
\end{tabular}
\end{table}

\begin{table}
\caption{Orbital parameters for HD 110014 b and HD 110014 c.}             
\label{HD110014_table2}    
\centering                          
\begin{tabular}{l l l}        
\hline\hline                 
& b & c \\
\hline
$P$ (day)&882.6 $\pm$ 21.5 & 130.0 $\pm$ 0.9 \\   
$T_0$ (JD-2450000)& 3739.4 $\pm$ 50.2 & 4031.3 $\pm$ 3.78\\
$e$ & 0.26 $\pm$ 0.1 & 0.44 $\pm$ 0.2\\
$\omega$ (deg)& 47.72 $\pm$ 15.5 & 307.6 $\pm$ 19.9\\
$M_p\sin i$ ($M_{\mbox{\scriptsize{Jup}}}$)& 10.7 $\pm$ 1.0 & 3.1 $\pm$ 0.4 \\
$a$ (AU)& 2.31 $\pm$ 0.04 & 0.64 $\pm$ 0.003\\
\hline                                  
\end{tabular}
\end{table} 

\begin{figure}
\centering
\includegraphics[width=8.8cm, height=6cm]{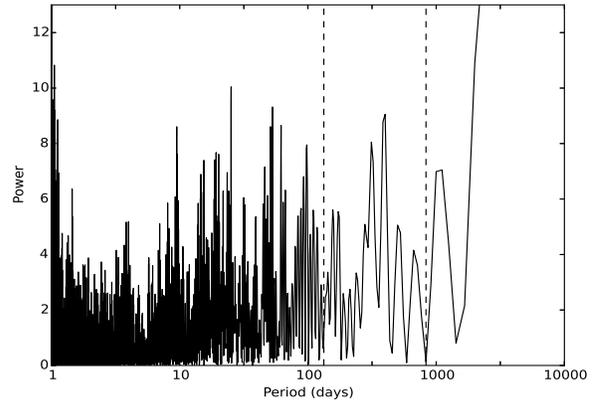}
\caption{Lomb-Scargle periodogram of the sampling of the data for HD 110014. The vertical dashed lines represent the location of the 833-days period and the 133-days period found in the periodograms of Figures~\ref{HD110014_periodogram} and~\ref{HD110014_periodogram_residuals}.} 
\label{HD110014_residuals_sampling}
\end{figure}

We performed a F-test between the one-planet solution and the two-planet solution and found that they are significally different (probability of $\sim$0\% that they are both statistically similar).
We also looked at the periodogram of the sampling (Figure~\ref{HD110014_residuals_sampling}) and we find that there is no peak corresponding to the 133-days period. This helps us to rule out the possibility that the signal we see is due to the sampling of the data.

In M09 a periodicity of 130-days was also detected in the residuals of their fit. They performed several tests in order to discard the possibility that the signal is intrinsic to the star. They could not find any correlation between those tests and the measured velocities, nor with the fit residuals, yet, they could not find enough evidence to be sure that the signal is produced by the presence of a second companion. The RMS they obtain after fitting the orbital solution of the second planet is 32.9 m s$^{-1}$, significantly higher than the 19.4 m s$^{-1}$ RMS we got with our fit.

The 130-days period could be caused, not only by a planetary companion, but also by a cool spot on the surface of the star. It has been shown that RV signals caused by rotational modulation of star spots can mimic Doppler signals, causing possible misinterpretation of the measurements (e.g. \citealt{setiawan08}; \citealt{huelamo08}), particularly when the star is at low inclination since some activity indicators, like the bisector inverse slope (BIS), are insensitive in this regime (\citealt{desort07}). In order to test this possibility, we first computed the spot filling factor from \citet{hatzes02}. This is the percentage of the surface of the star that a spot should have to cover in order to produce a RV signal with the same amplitude as the one measured. We used $V=v\sin i=1.1$ km s$^{-1}$, derived in \citet{deMedeiros14}, as the rotational velocity of the star, and $K_2=83.5$ m s$^{-1}$ as the RV amplitude produced by the 130-days signal. We obtained a filling factor of 13.8\%. In section \ref{photometry}, we discuss the analysis of available photometry for this star, to test if we can find signals that could be due to spots with this level of spot coverage.

\subsection{HD 122430}

HD 122430 is a giant star (K2III) with metallicity of [Fe/H] = -0.09 dex and a mass of 1.68 $M_{\odot}$ (data from M13). A planet orbiting this star was reported in \citet{setiawan_HD122430}, using data from FEROS and CORALIE. The orbital parameters correspond to a planet with a period of 344 days, eccentricity of 0.68, and minimum mass $M_p\sin i$ = 3.71-6.04 $M_{\mbox{\scriptsize{Jup}}}$ (for $M_{\star}$=1.2-2.5 $M_{\odot}$). The RMS of this solution is reported to be 27.2 m s$^{-1}$.

\begin{figure}
\centering
\includegraphics[width=8.8cm, height=6cm]{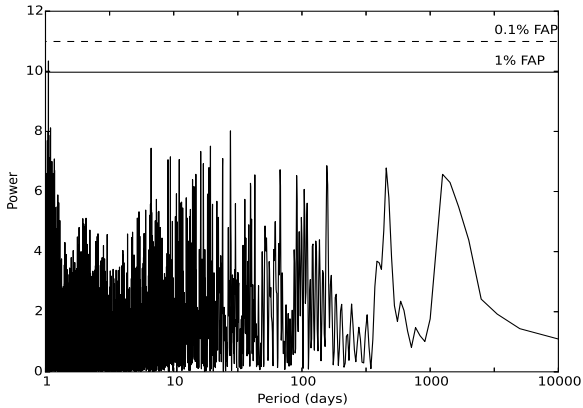}
\caption{Lomb-Scargle periodogram of HD 122430. The two horizontal lines represent the 1\% (solid line) and 0.1\% (dashed line) FAP.} 
\label{HD122430_periodogram}
\end{figure}

\begin{figure}
\centering
\includegraphics[width=8.8cm]{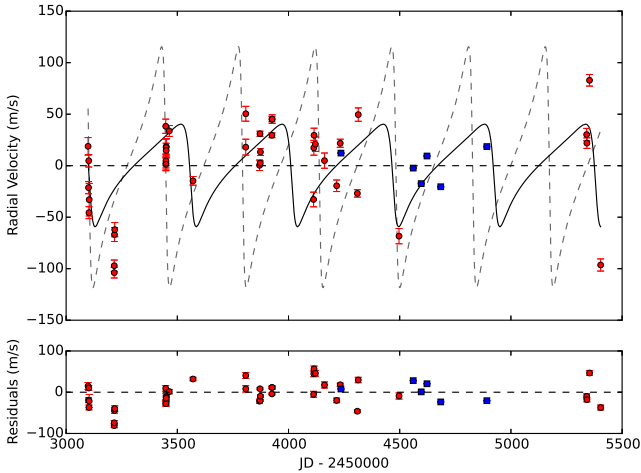}
\caption{{\it Top panel}: Radial velocity measurements for HD 122430 from FEROS (red circles) and HARPS (blue squares). The back solid line corresponds to the RV curve found in this paper and has a RMS of 29.3 m s$^{-1}$. The grey dashed line is the solution found in \citet{setiawan_HD122430} and gives a RMS of 63.4 m s$^{-1}$. {\it Bottom planel}: Residuals from the fit.}
\label{HD122430}
\end{figure}

We used spectra taken with FEROS from 2004 to 2010, totaling 42 data points. We also used 6 velocities computed with HARPS as a check on our solution. When trying to fit the values for the orbital parameters found by \citet[][hereafter S03]{setiawan_HD122430} we found a RMS of 63.4 m s$^{-1}$. From the periodogram (Figure~\ref{HD122430_periodogram}) we could not find any significant peak that would lead us to a starting period for a fit. We then let the console minimize the orbital parameters with starting values given by the orbital parameters found in S03. This leads us to a new solution with a period of 455.1 days, an eccentricity of 0.6, a minimum mass of 2.1 $M_{\mbox{\scriptsize{Jup}}}$ and a RMS of 29.3 m s$^{-1}$. The fit is shown in Figure~\ref{HD122430} as the black solid line, where the red circles and the blue squares represent the FEROS and HARPS data, respectively. Even though the RMS of S03 is still lower than the one obtained for HD 47536 and HD 110014 (one-planet fit), their periodograms showed at least one period with a significant peak close to the 1\% and 0.1\% FAP, which helped us prove that there was a signal within the RV data.  This is not the case for HD 122430, and after finding a RMS of 29 m s$^{-1}$ for our best solution, we cannot support the existence of this planet based on our data. Indeed, a mean noise model fit to the data returns an RMS of 40 m s$^{-1}$, only $\sim$ 10 m s$^{-1}$ larger than the one-planet solution.

In order to check if we missed the signal in the LS periodogram because of its high eccentricity, we also calculated the Generalized Lomb-Scargle Periodogram (GLS, \citealt{zechmeister09}). This periodogram is more suited to detecting signals arising from bodies on eccentric orbits ($e>0.5$). We found no peak above the noise level, meaning that the high eccentricity was not the only reason no peak was found in the LS periodogram. Therefore, if a planet does exist orbiting this star, we do not have the necessary data yet to make this claim.

\subsection{HD 70573}

The last star in our sample is HD 70573, a young lithium bearing star classified as a member of the Local Association (Pleiades moving group) by \citet{montes01}.  The age is therefore estimated to be between 78-125 Myr (only 3\%-6\% the age of the Sun, where moving groups are thought to be associations of young stars, see \citealt{murgas2013}). It has a spectral type of G1-1.5 V and a mass of 1.0 $M_{\odot}$, according to \citet[][hereafter S07]{setiawan_HD70573}. \citet*{henry95} measured photometric variations of HD 70573 and found a rotational period of 3.296 days. 

\begin{figure}
\centering
\includegraphics[width=8.8cm]{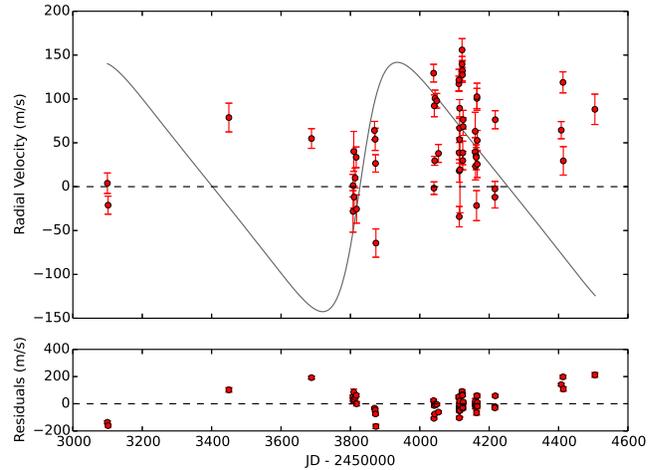}
\caption{{\it Top panel}: Radial velocities measured by FEROS for HD 70573 (red dots). The grey line correspond to the orbital solution found in S07. The RMS of this fit is 78.8 m s$^{-1}$. {\it Bottom panel}: Residuals from the fit. A long period variation can be seen in the residuals.}
\label{HD70573_setiawan}
\end{figure}

\begin{figure}
\centering
\includegraphics[width=8.8cm, height=6cm]{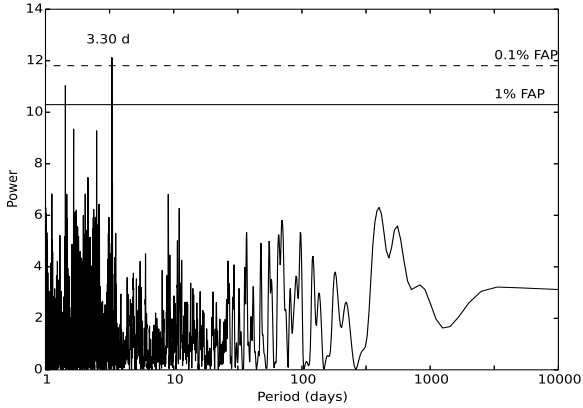}
\caption{LS periodogram of HD 70573. The two horizontal lines represent the 1\% (solid line) and 0.1\% (dashed line) FAP. Only one significant peak, above the 0.1\% FAP, is seen, corresponding to 3.296 days.} 
\label{HD70573_periodogram}
\end{figure}

\begin{figure}
\centering
\includegraphics[width=8.8cm]{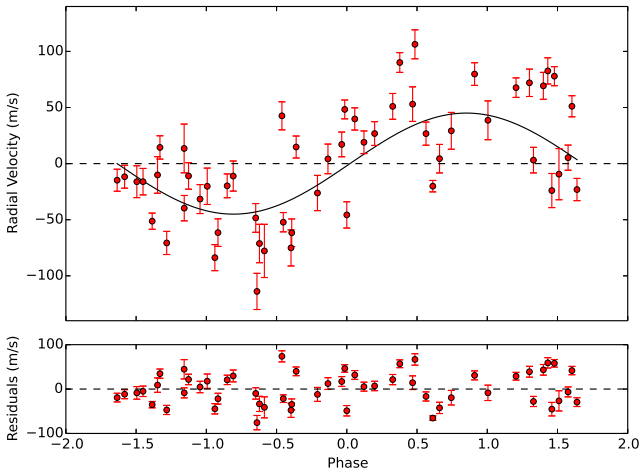}
\caption{{\it Top panel}: Phase-folded RV curve for HD 70573 (red dots), and fit to the data (solid line).The RMS of this fit is 36.4 m s$^{-1}$. {\it Bottom panel}: Residuals from the fit.}
\label{HD70573}
\end{figure}

This star was studied in S07, in which they claimed the detection of a companion with a period of 852 days and a minimum mass ($M_p\sin i$) of 6.1 $M_{\mbox{\scriptsize{Jup}}}$. To verify this detection we analyzed data taken with FEROS between 2004 and 2008, corresponding to a total of 41 nights and 55 data points. Since this is a young star some parts of its spectrum contain very few lines or heavily blended lines (because of line broadening from rapid rotation). This leads to an innacurate correlation between the spectra and the template for some orders, and in turn in noisy results for the velocities. To deal with this, we had to modify the orders that we were considering to calculate the RVs. After visually  inspecting each order for different observations of this star, we decided to consider only from order 9 to 31, instead of using information from order 2 to 36 like we did for the above stars.

\begin{figure}
\centering
\includegraphics[width=8.8cm, height=6cm]{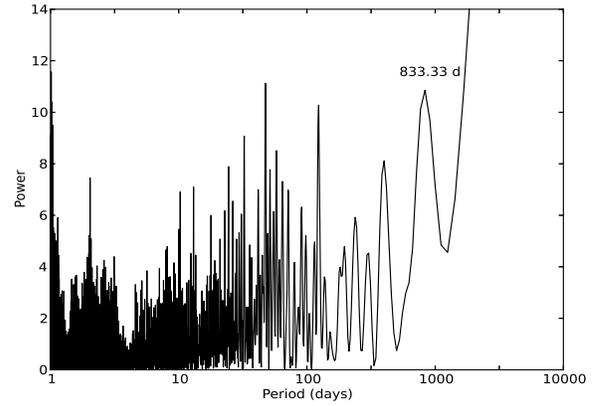}
\caption{LS periodogram of the sampling window for HD 70573. The peak corresponding to 833 days, close to the period of 852-days found by S07, is shown.} 
\label{HD70573_sampling}
\end{figure}

When we fit a companion with the orbital parameters as given by S07 to our data we arrived at a RMS of 78.78 m s$^{-1}$, much higher than the published RMS of 18.7 m s$^{-1}$. The fit is shown in Figure~\ref{HD70573_setiawan} as the grey solid line, and the red circles represent the FEROS RV's. The periodogram of the data (Figure~\ref{HD70573_periodogram}) does not show any significant peak at high periods, like it did in S07. Only one peak above the 0.1\% FAP can be seen, corresponding to a period of 3.296 days, the same as the rotational period of the star found in \citet{henry95}. We tried to fit a curve with a  period of 3.296 days to model the rotational effects of the star on the RVs.  The fit is shown in Figure~\ref{HD70573} and has an associated RMS of 36.4 m s$^{-1}$.
The periodogram of the sampling (Figure~\ref{HD70573_sampling}) shows that one of the peaks correspond to a 833-day period, close to the 852-days period proposed by S07. This indicates that the previously published orbital solution might be due to a sampling alias combined with noisy data.

\section{Activity}

\subsection{S-index and Bisector Velocity Span}

It has been shown that intrinsic stellar phenomena such as magnetic activity can produce a periodic radial velocity signal that can be misinterpretated as a substellar companion (\citealt{queloz01}). We performed three tests of our data to verify if this is the case for the stars we are studying. First, we measured the change in the flux from the Ca II HK lines over time by calculating the S-index for each epoch (the process is described in \citealt{jenkins08,jenkins11}). The second test we used was the bisector analysis (\citealt{bisector}), aimed at detecting asymmetries in the line profiles caused by intrinsic stellar phenomena. We measured the bisector velocity span (BVS) for each epoch, which essentially corresponds to the velocity difference between the bottom and top thirds of the cross correlation function (CCF, see \citealt{jones13}).  Finally, the third test we used to parameterize line asymmetries was analysis of the CCF full width at half maximum (FWHM).

Our aim was to test if these activity indicators were correlated with the RV at each epoch. To measure the level of the correlation we computed the Pearson rank correlation coefficient, which is a measure of the linear correlation between two variables, where 1 is a total positive correlation, 0 is no correlation, and -1 is total negative correlation. When computing the Pearson correlation coefficient all the data points have the same weight, meaning that one point lying very far away from the mean is going to affect the final result, even if it has a large uncertainty. That is why we only used the points that were closer than 2.5$\sigma$ from the mean value of the distribution. 

\begin{figure}
\centering
\includegraphics[width=8.8cm]{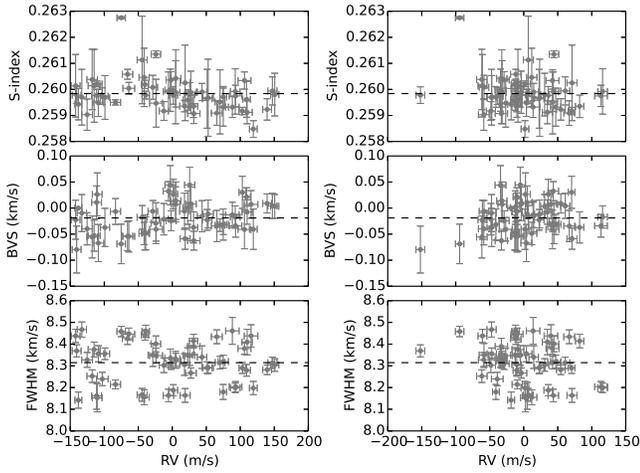}
\caption{S-index (upper panel), bisector velocity span (middle panel), and FWHM of the CCF (lower panel) versus the measured RV (left panel) and versus the residual RV for HD 47536, after fitting the one-planet solution. Only the values for the activity indices laying 2.5$\sigma$ from the mean value of the distribution were considered. The correlation coefficients for the plots in the left panels are $r$= -0.3, 0.3 and -0.02, and for the right panels are $r=-0.2,$ 0.24 and -0.23 for the S-index, BVS and FWHM versus the residual velocity, respectively. Horizontal dashed lines correspond to the mean of the distribution for each plot.}
\label{HD47536_activity}
\end{figure}

For most of the stars no correlation was found between the activity indicators and the velocities, with $r\sim 0$, where $r$ is the Pearson correlation coefficient. Plots of the indices vs RV for each star can be found in the Appendix. There were some cases that required further analysis. One of them is HD 47536. The RVs obtained for this star showed a lot of jitter, even after fitting a one-planet solution ($\sim$ 50 m s$^{-1}$), leaving us with a RMS of 51.7 m s$^{-1}$ which could not be significantly improved by adding a new component to the system. In the left panels of Figure~\ref{HD47536_activity} we show the activity indicators versus the measured RV, and in the right panels the same activity indicators but this time versus the residuals left over after fitting our one-planet solution. There are no significant correlations in any of the plots versus the RVs, with $r$ indices equal to -0.3, 0.3 and -0.02. 
From the $\chi^2$ test between the indices and a mean noise model fit, we found that the probability of these being fit by a horizontal line is $\sim$ 90\%. 
This means that it is unlikely that the velocities are a product of the stellar activity, and are likely produced by an external source. Also, we find no correlation in any of the plots versus the residual RV (right panel), which might indicate that the jitter in the residuals is not due to magnetic activity nor rotationally modulated spots on the star.
We computed the velocity oscillation amplitude for this star from the relations in \citet{kjledsen95}. We used $M=0.98$ $M_{\odot}$ from M13 and $L=175.19$ $L_{\odot}$ from \citet{anderson12}. We obtained $v_{\mbox{\footnotesize{osc}}}=(41.8 \pm 2.5)$ m s$^{-1}$. 
This is only 10 m s$^{-1}$ lower than our RMS of 51.7 m s$^{-1}$, meaning that the residuals from our 1-planet fit could be explained by the expected ''jitter'' for this giant star.

\begin{figure}
\centering
\includegraphics[width=8.8cm]{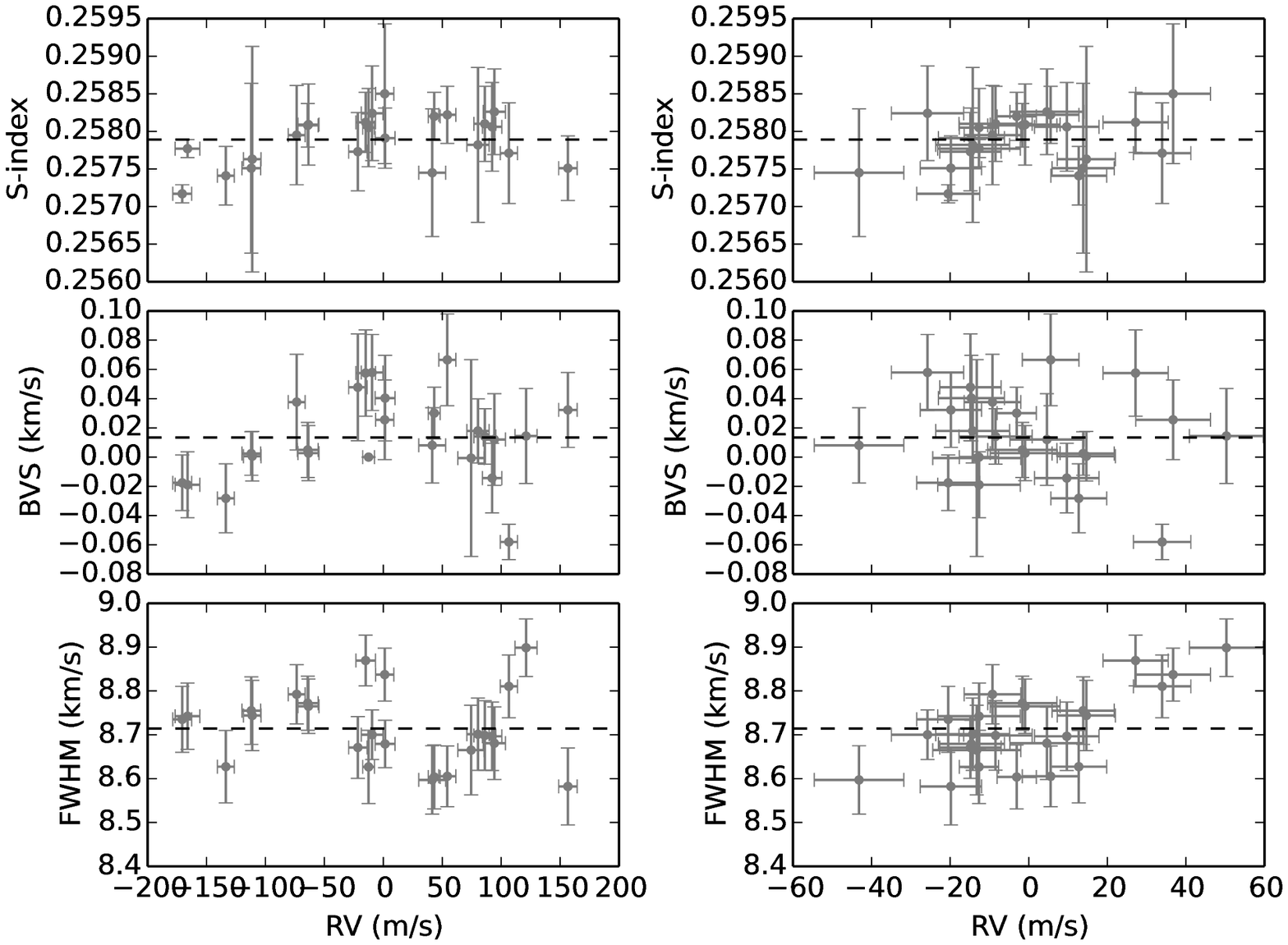}
\caption{S-index (upper panel), bisector velocity span (middle panel), and FWHM of the CCF (lower panel) versus the measured RV (left panel), and versus residual RV (right pannel) for HD 110014, after fitting the two-planet solution. Only the values for the activity indices laying 2.5$\sigma$ from the mean value of the distribution were considered. The correlation coefficients for the plots in the left panel are $r$=0.36, 0.19 and -0.17, and in the right panel are $r=0.3,$ -0.18 and 0.68 for the S-index, BVS and FWHM versus the velocity, respectively. Horizontal dashed lines correspond to the mean of the distribution for each plot.}
\label{HD110014_activity}
\end{figure}

In Figure~\ref{HD110014_activity} we can see the same plots as before, but this time for the star HD 110014. In the left panels we show the activity indices versus the measured RV, and in the right panels are the same activity indices but versus the residual velocity left after fitting the two-planet solution we are proposing in this paper. 
We found that the $r$ values for the activity indices and velocity in the left panels plots are $r$=0.36, 0.19 and -0.17 for the S-index, BVS and FWHM, respectively. This indicates that activity in the star is not the cause of the velocities we observe and supports our argument that they are produced by two planets orbiting the star. In the right panels, 
the $r$ indices are $r=0.3,$ -0.18 and 0.68 for the S-index, BVS and FWHM versus the residual velocity, respectively, suggesting that part of the jitter we were left with after fitting our two-planet solution may be caused by activity in the atmosphere of the star.

\begin{figure}
\centering
\includegraphics[width=8.8cm, height=6cm]{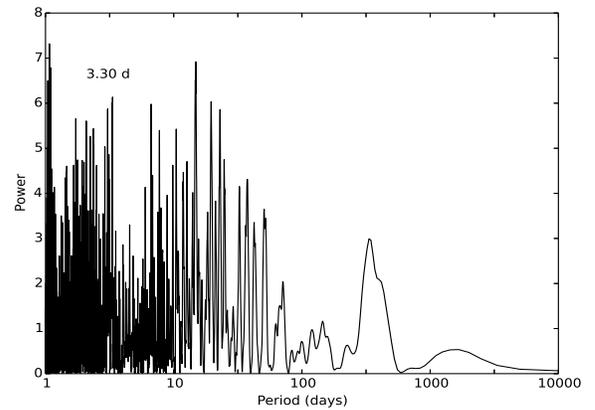}
\caption{LS periodogram of the bisector velocity span for HD 70573.} 
\label{HD70573_bvs_periodogram}
\end{figure}

\begin{figure}
\centering
\includegraphics[width=8.8cm]{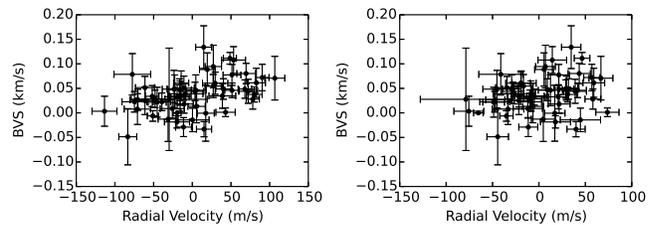}
\caption{BVS vs measured RV (left panel) and vs residual velocity after removing the rotational solution (right panel) for the star HD70573. The $r$ coefficients are 0.44 (left panel) and 0.26 (right panel).} 
\label{HD70573_bvs_rv}
\end{figure}

The last case is for HD 70573. One interesting result is that, when we studied the periodogram of the BVS (Figure~\ref{HD70573_bvs_periodogram}), we found a signal corresponding to the orbital period of the star of 3.296 days, just like we did with our RVs. Even though this peak is not nearly as significant as it is in Figure~\ref{HD70573_periodogram}, it is present and it supports our result that the observed velocities contain a signal that is due to features on the star modulated by its rotation. The S-index and FWHM of the CCF do not show correlation with the RV, with $r$ indices less than 0.2. However, for the case of the BVS (Figure~\ref{HD70573_bvs_rv}), we have a correlation coefficient of $r=0.44$ for this index vs the measured RVs, which is suggestive of correlation. The $r$ value decreases to $r=0.26$ when considering just the residual velocities after fitting the rotational period of the star. This strongly suggests that the rotation of the star is likely the cause of the variation seen in the bisector.

\subsection{Photometry}\label{photometry}

\begin{figure}
\centering
\includegraphics[width=8.8cm]{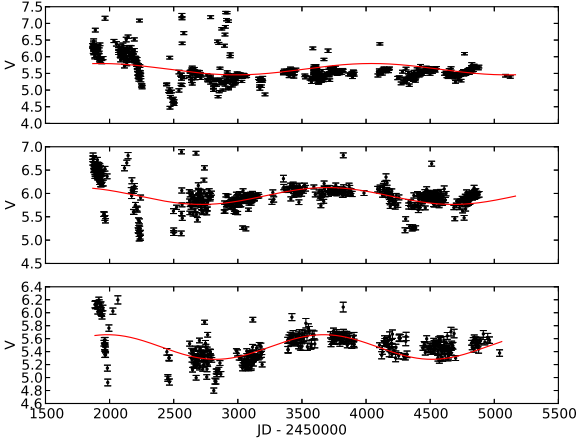}
\caption{V-magnitude variation for HD 11977 (upper panel), HD 47536 (middle panel) and HD 110014 (lower panel). In each plot a long-period variation was found (red lines), with $P=2114$ days for HD 11977, $P=1962$ days for HD 47536, and $P=1687$ days for HD 110014.} 
\label{Vmags}
\end{figure}

We looked for any period variation in the visual magnitude that could mimic the results we show in this work. We analyzed the $V$-band photometry from the All Sky Automated Survey (\citealt{photometry}) for the stars HD 11977, HD 47536 and HD 110014. The precision from this data is on the order of 1\%. The magnitude versus time are plotted in Figure~\ref{Vmags}. Superimposed on each plot are the long-period fits, with periods of $P=2114$, 1962 and 1687 days for HD 11977, HD 47536 and HD 110014, respectively. Because these are evolved stars, this long-period variation could be caused by magnetic cycles in the photosphere of the stars. These periods are not present in the velocities so we can conclude that this photometric variation is not the cause of the signals in the RV data. 

For HD 110014 we analyzed the photometry in order to search for any signals from cool spots on surface of the star that could cause the RV signal we found (see section \ref{hd110014}). We could not find any period above the noise level that could correspond to the periods of the planets detected. We also looked at the Hipparcos photometry (\citealt{perryman97}) for this star but we found no significant peak with a period longer than 2 days. The precision of the data for this star is $\sim$0.1\%. As previously mentioned, the filling factor of a spot on the surface would need to be 13.8\% to produce this level of RV signal, and therefore it is likely that this level of spot coverage would not be detectable in this dataset, meaning a spot origin for the signal can not be ruled out with any degree of certainty.  However, we do note that we studied the power of the periodogram peak as a function of time, as we did for HD 47536 in section~\ref{hd47536}. We found that, in general, the signal peak rising almost monotonically. This lends some weight to the planetary hypothesis for the second signal in the RV data of this star.

\section{Conclusions and Discussion}

We computed precise radial velocities using archival FEROS spectra for stars HD 11977, HD 47536, HD 70573, HD 110014 and HD 122430. From this data we can conclude the following:

\begin{itemize}

\item There is a companion around HD 11977 with a period of 621.6 days and minimum mass of 6.5 $M_{\mbox{\scriptsize{Jup}}}$.

\item We can confirm the presence of one companion around HD 47536, with a period of 434.9 days and minimum mass of 4.0 $M_{\mbox{\scriptsize{Jup}}}$. We cannot confirm the existence the second companion proposed by S08.

\item We confirm the existence of a companion around HD 110014, with a period of 882.6 days and minimum mass of 10.7 $M_{\mbox{\scriptsize{Jup}}}$. We also found evidence of a second companion around this star with a period of 130.0 days and minimum mass of 3.1 $M_{\mbox{\scriptsize{Jup}}}$. A more extensive photometric analysis has to be done to confirm that this signal corresponds to a second companion and not to a cold spot on the star's surface.

\item There is no signal significant enough to confirm the presence of any companion around HD 122430.

\item We found no evidence of the planet orbiting HD 70573 proposed in S07, nor evidence of any other companion. The only signal present corresponds to the rotational period of the star.

\item There is no correlation between the activity indices and the RVs for the stars HD 11977, HD 47536, HD 110014 and HD 122430, supporting our hypothesis that the observed velocity signals are Doppler signals. We find a small yet not significant correlation between the activity indices and the measured RVs for HD 70573, but it decreases after removing the rotation of the star. We find a higher level of correlation in the activity indices versus the residual velocities after removing the two-planet fit for HD 110014. This could explain the jitter in the residual RVs as a product of activity within the star.

\item We observe a periodic variation in the photometric data for three of the stars in our sample (HD 11977, HD 46536 and HD 110014), but with periods different than the ones of the companions we found. We conclude that the signals in HD 11977 and HD 47536 data cannot be explained by intrinsic photometric variation in the stars. For HD 110014 we could not find evidence of any photometric signal with periods equal to the one of the RV signals, supporting our argument of the existence of this second planet orbiting this star.

\end{itemize}

\begin{figure}
\centering
\includegraphics[width=7cm]{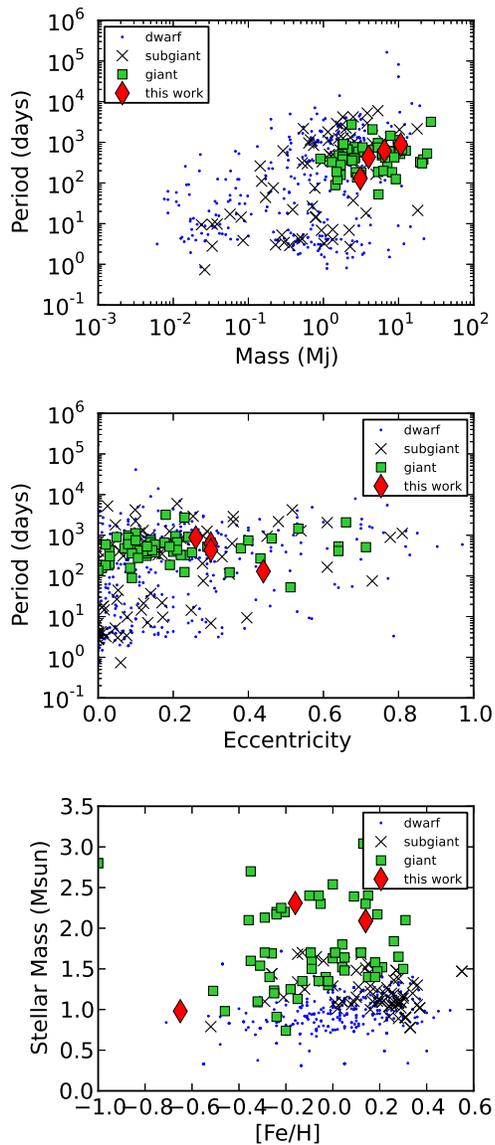}
\caption{Period versus mass (upper panel), period versus eccentricity (middle panel) and stellar mass versus stellar metallicity (bottom panel) for all confirmed planets with known values for the stellar type, mass and metallicity, and for known values of planetary mass, eccentricity and period. There is a total of 317 planets around dwarf stars, 76 around sub-giant stars, and 67 around giant stars (including the ones in this work).} 
\label{mult}
\end{figure}

Finally, in Figure~\ref{mult} we can see that the planets found in this work follow similar trends to the previously detected planets around giant stars. The upper and middle panels show that these planets are mostly clustered around periods from $\sim$ 100 to 1000 days, and with masses from $\sim$ 1 to 30 $M_{\mbox{\scriptsize{Jup}}}$. The bottom panel of Figure~\ref{mult} is similar to Figure 2 in \citet[][hereafter R15]{reffert15} which shows the distribution of metallicity and stellar mass for planets around giant stars. HD 110014 b and c fall in the region corresponding to a 20.8\% probability of the star having a planet, and HD 11977 b falls in the block with probability of 6.9\%. Both stars are in the range of masses for giant stars for which we can expect the highest number of planets ($\sim$2 $M_{\odot}$ according Figure 2 and 5 of R15). HD 110014 is also the star with the highest metallicity in our sample, and in turn is the one with the most currently detected planets (two found in this work). This follows the notion that the higher the metallicity of a star, the higher the probability of finding a planet orbiting them. This result was discussed in R15 and mimics the trend and discoveries of planets orbiting the most metal-rich lower-mass stars (\citealp{fischer05}; \citealp{sousa11}; \citealp{jenkins13}). 

One special case is that of HD 47536 b. This planet is orbiting a very metal poor giant star ([Fe/H]=-0.65 dex), making HD 47536 the second most metal-poor star known to host a planet (the most metal-poor star was thought to be BD+20 2457, with [Fe/H]=-1.0 dex, but it was recently shown to be an unstable system; see \citealp{horner14}). Actually, the probability of finding a planet around such a metal poor star is almost 0\% (according to Figure 4 and 7 of R14). The confirmation of a planet orbiting HD 47536 is a very interesting case to consider in the study of the formation of these systems.

\section*{Acknowledgments}

MGS acknowledges support from CONICYT-PCHA/Doctorado Nacional/2014-21141037. JSJ acknowledges support from the CATA-Basal grant (PB06, Conicyt). M.J. acknowledges financial support from Fondecyt project \#3140607 and FONDEF project CA13I10203. We acknowledge the very helpful comments from the anonymous referee. Based on data obtained from the ESO Science Archive Facility under request number maritsoto 157382, 157389, 157391, 157392, 161298, 161300, 161301, 161302 and 161303.

\bibliographystyle{mn2e}
\bibliography{refs}{}

\newpage
\appendix

\section{RV measurements}

\begin{table}
\caption{RV measurements for HD 11977.}       
\centering                          
\begin{tabular}{c c c c c c}    
\hline
\multicolumn{6}{c}{FEROS}\\
\hline
\hline
JD - & RV & Error & JD - & RV & Error\\
2450000 & (m/s) & (m/s) & 2450000 & (m/s) & (m/s)\\
\hline
3100.480 & -54.5 & 7.5 & 3585.911 &  4.9 & 8.2 \\
3101.488 & -32.4 & 8.6 & 3586.743 & 15.2 & 8.7 \\
3102.481 & -82.6 & 6.9 & 3586.903 & -0.4 & 7.9 \\
3102.486 & -74.4 & 7.1 & 3590.726 & -5.1 & 9.0 \\
3216.910 & -54.8 & 7.5 & 3591.767 &  0.4 & 8.3 \\
3217.842 & -66.7 & 11.1 & 3591.931 &  1.4 & 7.2 \\
3218.933 & -59.6 & 8.1 & 3592.733 & 10.0 & 9.9 \\
3331.581 & -12.9 & 5.9 & 3592.933 &  3.6 & 9.5 \\
3332.556 & -2.8 & 5.8 & 3593.725 & -11.4 & 9.9 \\
3334.622 & 11.6 & 6.5 & 3593.929 & -8.4 & 9.0 \\
3335.688 &  6.3 & 5.4 & 3594.715 & -8.0 & 9.0 \\
3336.691 & 16.3 & 5.2 & 3594.926 & -2.6 & 7.9 \\
3338.709 & -12.8 & 6.0 & 3595.853 &  2.5 & 14.2 \\
3448.479 & 71.6 & 7.4 & 3932.943 & -20.7 & 7.2 \\
3449.481 & 107.5 & 4.1 & 4041.589 & 44.3 & 6.3 \\
3449.484 & 114.9 & 4.8 & 4112.543 & 115.1 & 5.1 \\
3581.730 & 11.1 & 9.4 & 4113.550 & 116.7 & 3.4 \\
3581.925 & 10.6 & 8.3 & 4115.552 & 105.9 & 2.4 \\
3582.737 & 25.1 & 9.7 & 4313.859 & -65.9 & 16.1 \\
3582.929 &  8.0 & 9.0 & 4315.791 & -85.3 & 8.7 \\
3583.726 & 24.2 & 9.7 & 4316.850 & -78.8 & 7.7 \\
3583.945 & 23.5 & 9.0 & 4316.927 & -45.1 & 7.2 \\
3584.705 & 25.5 & 8.9 & 4317.857 & -90.2 & 14.9 \\
3585.720 & 11.1 & 9.3 & 4319.871 & -66.3 & 8.1 \\
\hline  
\multicolumn{6}{c}{HARPS}\\
\hline
\hline
JD - & RV & Error & JD - & RV & Error\\
2450000 & (m/s) & (m/s) & 2450000 & (m/s) & (m/s)\\
\hline  
3334.620 & 49.2 & 0.2 & 4787.602 & 120.8 & 0.3 \\
4045.592 & 121.1 & 0.6 & 4891.528 & -10.6 & 0.4 \\
4370.721 & -30.1 & 0.3 & 5083.811 & -1.5 & 0.3 \\
4371.671 & -7.5 & 0.4 & 5084.748 & -2.9 & 0.2 \\
4622.915 & 55.1 & 0.4 & 5085.754 & -8.5 & 0.2 \\
4684.748 & 128.8 & 0.3 & 5086.730 & -2.1 & 0.4 \\
4785.641 & 131.2 & 0.6 & & & \\
\hline  
\multicolumn{6}{c}{CHIRON}\\
\hline
\hline
JD - & RV & Error & JD - & RV & Error\\
2450000 & (m/s) & (m/s) & 2450000 & (m/s) & (m/s)\\
\hline  
6906.820 & -17.7 & 5.4 & 6965.553 & -6.1 & 4.6 \\
6919.634 & -16.1 & 4.0 & 6988.552 & 10.6 & 5.2 \\
6921.717 & -22.6 & 4.7 & 6993.525 & 19.6 & 5.0 \\
6964.740 & -3.5 & 4.6 & 7013.531 & 35.7 & 5.2 \\
\hline               
\end{tabular}
\end{table}

\begin{table}
\caption{RV measurements for HD 47536.}    
\centering                          
\begin{tabular}{c c c c c c}    
\hline
\multicolumn{6}{c}{FEROS}\\
\hline
\hline
JD - & RV & Error & JD - & RV & Error\\
2450000 & (m/s) & (m/s) & 2450000 & (m/s) & (m/s)\\
\hline
3454.506 & -65.8 & 7.7 & 4217.484 & -99.0 & 6.9 \\
3454.509 & -63.9 & 8.7 & 4231.459 & -125.2 & 7.9 \\
3460.566 & -39.0 & 5.5 & 4233.471 & -24.1 & 7.0 \\
3460.569 & -40.2 & 5.3 & 4402.754 & 150.7 & 6.0 \\
3685.756 & 27.3 & 8.0 & 4407.858 & 109.3 & 5.6 \\
3687.752 & 52.8 & 7.6 & 4492.534 & 115.9 & 10.0 \\
3688.716 & 51.3 & 7.1 & 4493.533 & 110.1 & 7.5 \\
3689.708 & 75.0 & 7.6 & 4494.540 & 88.4 & 9.9 \\
3689.857 & 70.3 & 7.2 & 4509.602 & 31.7 & 9.1 \\
3690.663 & -1.7 & 7.4 & 4510.579 & 106.2 & 9.5 \\
3809.589 & -142.1 & 6.5 & 4572.613 & -75.1 & 6.2 \\
3813.559 & -133.0 & 6.8 & 4582.500 & -140.6 & 6.5 \\
3869.512 & -108.7 & 8.1 & 4780.881 & 103.1 & 4.9 \\
3873.472 & -114.9 & 8.8 & 5109.833 & -83.5 & 7.5 \\
4040.759 & 147.3 & 6.7 & 5167.827 & -138.1 & 5.1 \\
4042.774 & 139.5 & 6.4 & 5201.738 &  1.5 & 6.4 \\
4112.653 &  5.1 & 7.6 & 5255.591 & 74.9 & 5.0 \\
4113.636 &  4.9 & 6.2 & 5258.694 & 119.2 & 5.7 \\
4114.664 & -12.4 & 6.7 & 5491.768 & 93.1 & 9.3 \\
4115.689 & 22.7 & 7.0 & 5494.891 & 92.6 & 7.1 \\
4120.765 & -2.8 & 8.0 & 5521.825 & -5.6 & 7.7 \\
4121.710 & 25.6 & 7.5 & 5523.764 & -44.3 & 7.4 \\
4122.790 & -28.6 & 7.0 & 5524.789 & 18.9 & 7.4 \\
4123.791 & 43.8 & 6.9 & 5525.881 & -41.4 & 7.5 \\
4124.815 & 18.0 & 7.1 & 5583.808 & -110.4 & 7.2 \\
4157.525 & -23.8 & 9.5 & 5588.777 & -111.3 & 7.2 \\
4162.558 & 65.5 & 8.0 & 5630.650 & -102.9 & 8.6 \\
4216.484 & 31.3 & 6.1 & 5632.546 & -117.8 & 7.7 \\
\hline  
\multicolumn{6}{c}{HARPS}\\
\hline
\hline
JD - & RV & Error & JD - & RV & Error\\
2450000 & (m/s) & (m/s) & 2450000 & (m/s) & (m/s)\\
\hline  
4045.814 & 18.4 & 0.3 & 4622.442 & -93.4 & 0.4 \\
4085.848 & 111.1 & 0.5 & 4623.462 & -25.5 & 0.7 \\
4188.594 & -70.4 & 1.1 & 4684.925 & -18.0 & 0.3 \\
4236.489 & -94.4 & 1.3 & 4685.925 &  7.4 & 0.4 \\
4370.863 & 84.4 & 0.3 & 4786.762 & 49.4 & 0.4 \\
4560.593 & -48.2 & 0.3 & 4788.863 & 39.1 & 0.3 \\
4561.578 & -54.1 & 0.3 & 4877.701 & 135.4 & 0.3 \\
4594.467 & -79.4 & 0.3 & 4891.645 & 66.1 & 0.3 \\
4596.469 & -101.1 & 0.5 & 4893.659 & 98.7 & 0.4 \\
\hline  
\multicolumn{6}{c}{CORALIE}\\
\hline
\hline
JD - & RV & Error & JD - & RV & Error\\
2450000 & (m/s) & (m/s) & 2450000 & (m/s) & (m/s)\\
\hline
2550.388 & -48.5 & 15.0 & 2719.400 & 240.6 & 10.0 \\
2622.090 & 76.7 & 10.2 & 2999.379 & 10.5 & 10.0 \\
2642.576 & 178.0 & 5.0 & 3001.086 & 29.7 & 10.0 \\
2669.891 & 140.9 & 15.0 & 3047.180 & 113.0 & 10.0 \\
2683.549 & 255.8 & 10.0 & 3052.302 & 130.2 & 10.0 \\
2717.693 & 192.4 & 10.0 & 3782.980 & -118.3 & 10.0 \\
\hline 
\multicolumn{6}{c}{CHIRON}\\
\hline
\hline
JD - & RV & Error & JD - & RV & Error\\
2450000 & (m/s) & (m/s) & 2450000 & (m/s) & (m/s)\\
\hline  
6953.723 & 13.8 & 6.8 & 6979.718 & -18.5 & 6.3 \\
6959.719 & -105.2 & 6.8 & 6987.636 & -36.6 & 5.8 \\
6967.768 & 48.5 & 6.2 & 7012.707 & 60.7 & 6.7 \\
6977.711 & 37.2 & 6.9 & & & \\
\hline               
\end{tabular}
\end{table}

\begin{table}
\caption{RV measurements for HD 110014.}  
\centering                          
\begin{tabular}{c c c c c c}    
\hline
\multicolumn{6}{c}{FEROS}\\
\hline
\hline
JD - & RV & Error & JD - & RV & Error\\
2450000 & (m/s) & (m/s) & 2450000 & (m/s) & (m/s)\\
\hline
4112.805 & -166.6 & 10.5 & 4316.480 & 91.8 & 8.2 \\
4113.784 & -171.2 & 8.0 & 4492.899 & 120.5 & 9.4 \\
4120.857 & -111.7 & 7.4 & 4495.852 & 106.0 & 7.3 \\
4120.861 & -112.5 & 7.9 & 5260.801 & 53.7 & 7.2 \\
4159.903 & -15.0 & 8.3 & 5262.783 & 42.7 & 5.0 \\
4161.809 & -5.6 & 9.5 & 5309.667 & 40.9 & 11.4 \\
4216.540 & -74.1 & 7.2 & 5356.570 & 156.1 & 7.8 \\
4217.765 & -64.0 & 8.9 & 5586.890 & -13.0 & 5.0 \\
4217.770 & -64.7 & 8.8 & 5615.805 & -10.1 & 9.2 \\
4309.474 & 79.8 & 9.4 & 5615.895 &  1.0 & 8.4 \\
4309.479 & 85.6 & 9.3 & 5630.733 & -22.2 & 7.8 \\
4312.487 & 93.6 & 9.4 & 5672.706 & -134.2 & 7.1 \\
4313.560 & 73.9 & 11.2 & & & \\
\hline  
\multicolumn{6}{c}{HARPS}\\
\hline
\hline
JD - & RV & Error & JD - & RV & Error\\
2450000 & (m/s) & (m/s) & 2450000 & (m/s) & (m/s)\\
\hline  
4235.759 & 38.1 & 0.3 & 4891.766 & -12.2 & 0.2 \\
4560.834 & 364.5 & 0.3 & 4931.641 & 20.2 & 0.2 \\
4594.768 & 289.4 & 0.4 & 5918.873 & -19.2 & 0.2 \\
4595.565 & 284.6 & 0.3 & 5934.864 & -28.9 & 0.2 \\
4596.735 & 287.5 & 0.4 & 5935.844 & -15.5 & 0.2 \\
4622.599 & 189.3 & 0.3 & 5936.861 & -16.4 & 0.2 \\
4623.663 & 204.5 & 0.4 & 5937.861 & -15.6 & 0.2 \\
4684.471 & 272.4 & 0.2 & 5938.860 & -37.8 & 0.2 \\
4685.514 & 247.0 & 0.3 & & & \\
\hline              
\end{tabular}
\end{table}

\begin{table}
\caption{RV measurements for HD 122430.} 
\centering                          
\begin{tabular}{c c c c c c}    
\hline
\multicolumn{6}{c}{FEROS}\\
\hline
\hline
JD - & RV & Error & JD - & RV & Error\\
2450000 & (m/s) & (m/s) & 2450000 & (m/s) & (m/s)\\
\hline
3098.776 & 18.8 & 8.5 & 3871.789 & 31.0 & 2.5 \\
3100.692 & -21.3 & 5.9 & 3872.753 &  2.5 & 2.1 \\
3101.783 &  4.8 & 6.1 & 3873.746 & 13.3 & 3.4 \\
3102.783 & -45.6 & 5.8 & 3873.771 & 19.4 & 3.0 \\
3103.692 & -33.0 & 15.8 & 3925.495 & 45.1 & 2.5 \\
3216.473 & -103.9 & 5.1 & 3925.601 & 29.5 & 2.3 \\
3216.479 & -97.1 & 5.5 & 3926.501 & 45.2 & 4.2 \\
3217.537 & -66.8 & 6.8 & 4112.821 & -32.8 & 7.1 \\
3218.561 & -62.1 & 6.9 & 4113.842 & 17.4 & 7.2 \\
3447.867 & 38.3 & 6.9 & 4114.843 & 29.5 & 6.9 \\
3448.836 &  2.2 & 6.4 & 4120.864 & 20.9 & 7.0 \\
3448.840 &  4.6 & 6.8 & 4161.812 &  4.9 & 7.4 \\
3448.845 &  1.3 & 6.2 & 4216.545 & -19.5 & 5.6 \\
3449.878 & 18.6 & 7.8 & 4232.597 & 21.8 & 3.9 \\
3449.882 & 17.7 & 8.1 & 4309.610 & -26.9 & 3.6 \\
3449.886 & 14.7 & 7.1 & 4313.621 & 49.6 & 6.5 \\
3463.889 & 33.7 & 5.3 & 4496.892 & -68.4 & 7.4 \\
3570.596 & -14.7 & 4.2 & 5340.713 & 30.2 & 6.0 \\
3807.795 & 18.0 & 7.7 & 5341.695 & 21.9 & 5.1 \\
3807.918 & 50.4 & 7.2 & 5353.529 & 82.9 & 5.6 \\
3869.794 &  0.4 & 5.1 & 5403.589 & -96.4 & 5.9 \\
\hline  
\multicolumn{6}{c}{HARPS}\\
\hline
\hline
JD - & RV & Error & JD - & RV & Error\\
2450000 & (m/s) & (m/s) & 2450000 & (m/s) & (m/s)\\
\hline  
4235.772 & 12.2 & 0.3 & 4622.627 &  9.4 & 0.3 \\
4560.852 & -2.3 & 0.3 & 4684.490 & -20.4 & 0.3 \\
4596.737 & -17.4 & 0.4 & 4891.774 & 18.6 & 0.3 \\
\hline              
\end{tabular}
\end{table}

\begin{table}
\caption{RV measurements for HD 70573.}   
\centering                          
\begin{tabular}{c c c c c c}    
\hline
\multicolumn{6}{c}{FEROS}\\
\hline
\hline
JD - & RV & Error & JD - & RV & Error\\
2450000 & (m/s) & (m/s) & 2450000 & (m/s) & (m/s)\\
\hline
3099.524 & -45.7 & 11.7 & 4114.743 & 17.1 & 11.0 \\
3101.539 & -70.7 & 10.3 & 4114.835 & 39.8 & 9.9 \\
3449.674 & 29.2 & 16.4 & 4115.721 & -30.3 & 49.1 \\
3687.838 &  5.3 & 11.2 & 4121.748 & 90.1 & 8.8 \\
3807.576 & -48.4 & 12.8 & 4121.856 & 106.3 & 12.9 \\
3807.638 & -77.8 & 23.7 & 4122.802 & 82.6 & 11.7 \\
3809.735 & -9.4 & 22.8 & 4122.849 & 77.9 & 8.5 \\
3810.604 & -61.5 & 12.3 & 4123.816 & -19.8 & 10.6 \\
3813.660 & -39.7 & 11.3 & 4123.859 & -11.1 & 13.4 \\
3816.663 & -16.1 & 11.8 & 4124.790 & 19.0 & 10.1 \\
3817.716 & -75.1 & 16.1 & 4124.866 & 26.8 & 10.6 \\
3869.524 & 14.4 & 10.4 & 4159.579 & -10.1 & 16.3 \\
3871.514 &  4.4 & 12.7 & 4159.769 & 13.5 & 21.7 \\
3872.493 & -23.1 & 9.9 & 4160.718 & -26.2 & 15.7 \\
3873.511 & -113.9 & 16.1 & 4162.588 & -14.8 & 9.8 \\
4039.874 & 79.8 & 10.1 & 4162.728 & -16.1 & 14.2 \\
4040.875 & -51.3 & 7.2 & 4163.601 & -71.2 & 17.1 \\
4041.799 & 42.6 & 12.5 & 4164.550 & 51.0 & 11.5 \\
4042.873 & -20.1 & 5.1 & 4164.691 & 53.0 & 15.4 \\
4043.863 & 51.1 & 9.4 & 4165.552 &  3.1 & 11.4 \\
4048.838 & 48.3 & 8.4 & 4165.683 & -24.0 & 15.2 \\
4053.864 & -11.7 & 10.2 & 4216.511 & -52.2 & 8.6 \\
4112.688 & 67.7 & 8.7 & 4216.571 & -61.6 & 12.3 \\
4112.783 & 72.0 & 12.2 & 4217.527 & 26.7 & 10.3 \\
4113.652 & -11.0 & 11.8 & 4407.787 & 14.8 & 9.8 \\
4113.734 & -31.6 & 12.9 & 4412.843 & 69.3 & 12.0 \\
4113.839 & -83.8 & 11.6 & 4413.747 & -20.2 & 16.2 \\
4114.645 &  4.2 & 13.2 & 4504.745 & 38.6 & 17.3 \\
\hline               
\end{tabular}
\end{table}

\section{Activity indices}

\begin{figure}
\centering
\includegraphics[width=8.8cm]{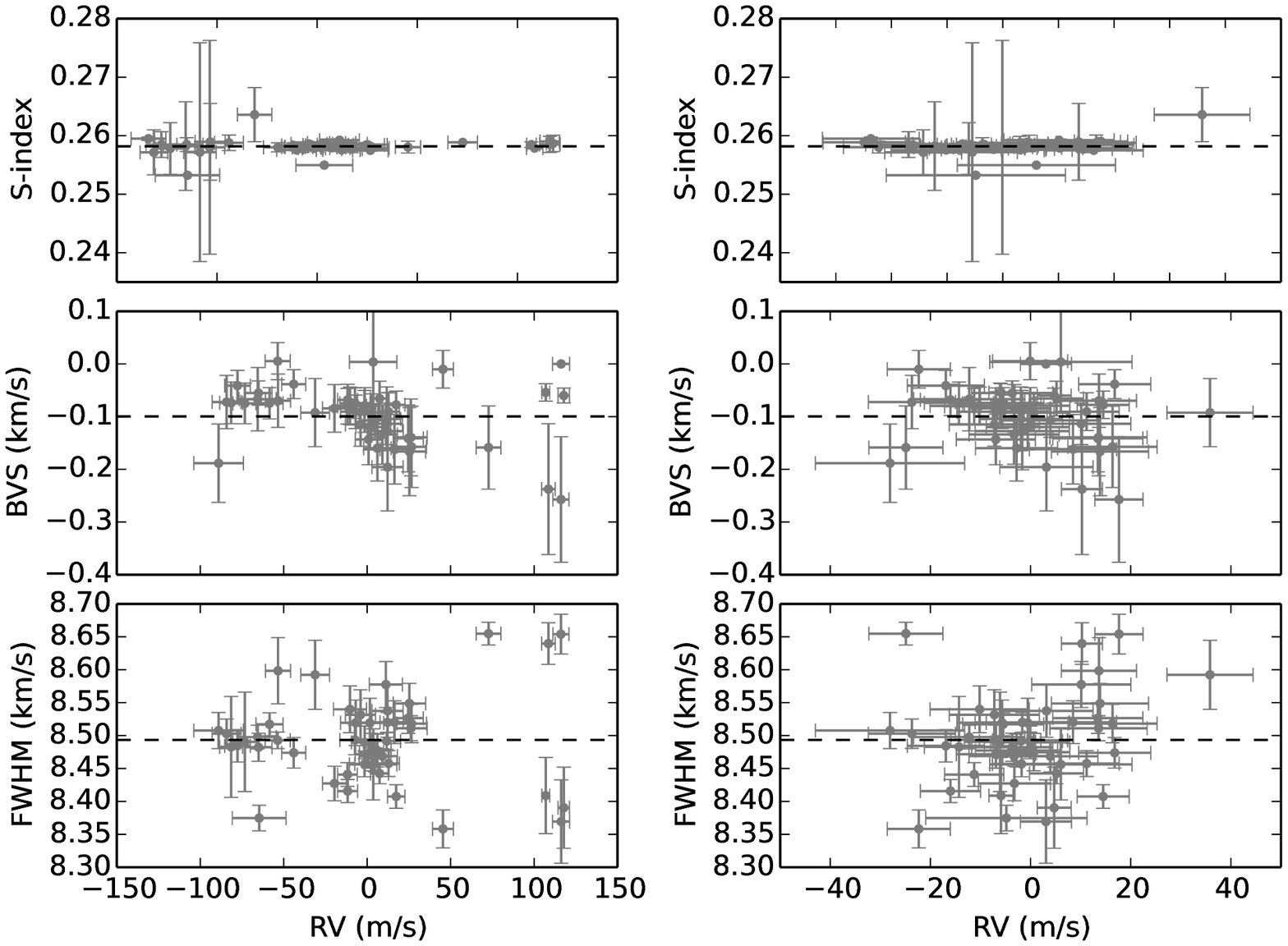}
\caption{S-index (upper panel), bisector velocity span (middle panel) and FWHM of the CCF (lower panel) versus the measured RV (left panel) and versus the residual RV for HD 11977, after fitting the 1-planet solution. Only the values for the activity indices laying 2.5$\sigma$ from the mean value of the distribution were considered. The correlation coefficients for the plots in the left panels are $r$= 0.11, -0.29 and 0.05, and for the right panels are $r=0.3,$ -0.19 and 0.25 for the S-index, BVS and FWHM versus the residual velocity, respectively. Horizontal dashed lines correspond to the mean of the distribution for each plot.}
\label{HD11977_activity}
\end{figure}

\begin{figure}
\centering
\includegraphics[width=8.8cm]{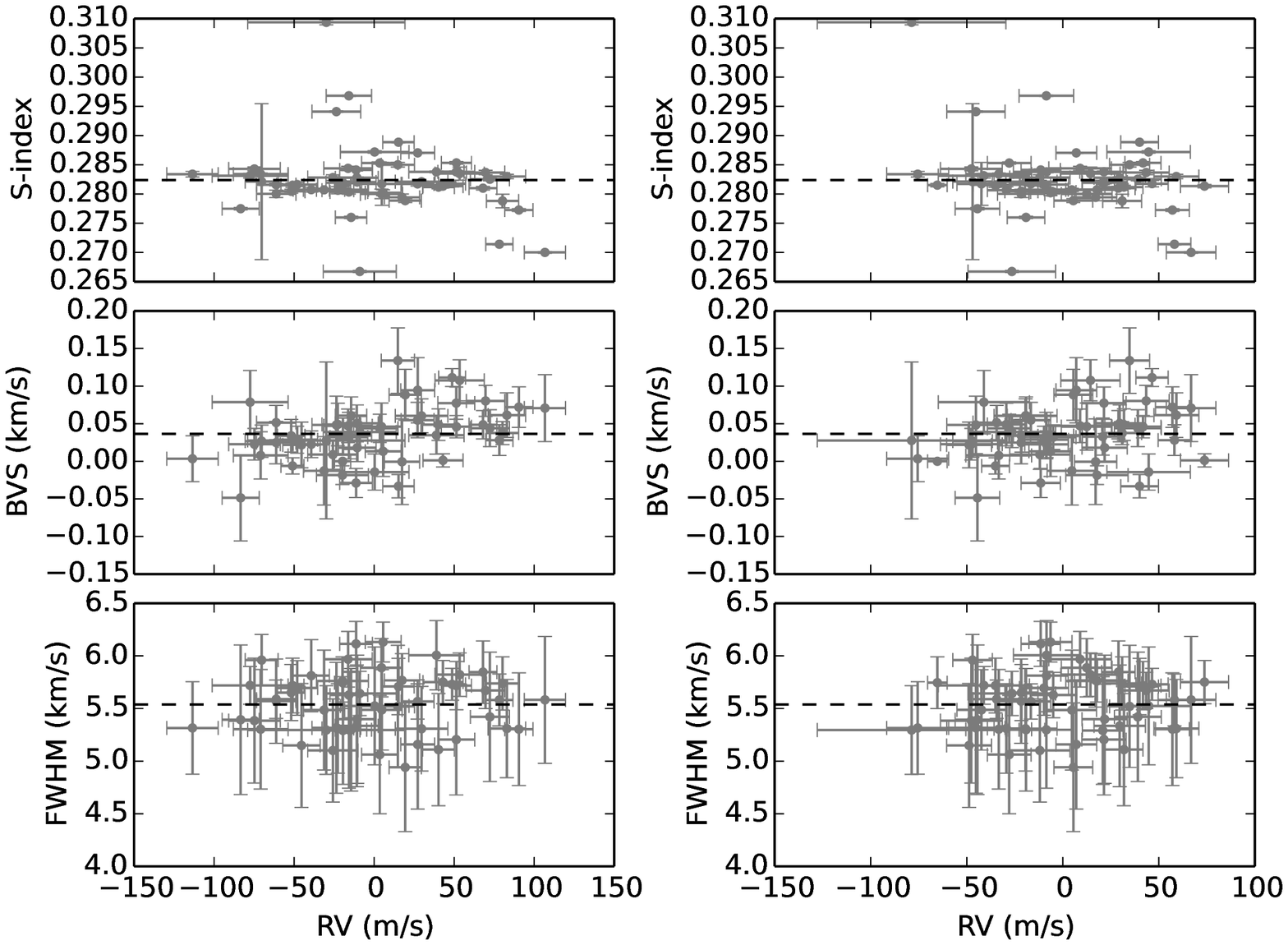}
\caption{S-index (upper panel), bisector velocity span (middle panel) and FWHM of the CCF (lower panel) versus the measured RV (left panel) and versus the residual RV for HD 70573, after removing the rotation of the star. Only the values for the activity indices laying 2.5$\sigma$ from the mean value of the distribution were considered. The correlation coefficients for the plots in the left panels are $r$= -0.19, 0.44 and 0.05, and for the right panels are $r=-0.28,$ 0.26 and 0.11 for the S-index, BVS and FWHM versus the residual velocity, respectively. Horizontal dashed lines correspond to the mean of the distribution for each plot.}
\label{HD70573_activity}
\end{figure}

\begin{figure}
\centering
\includegraphics[width=8.8cm]{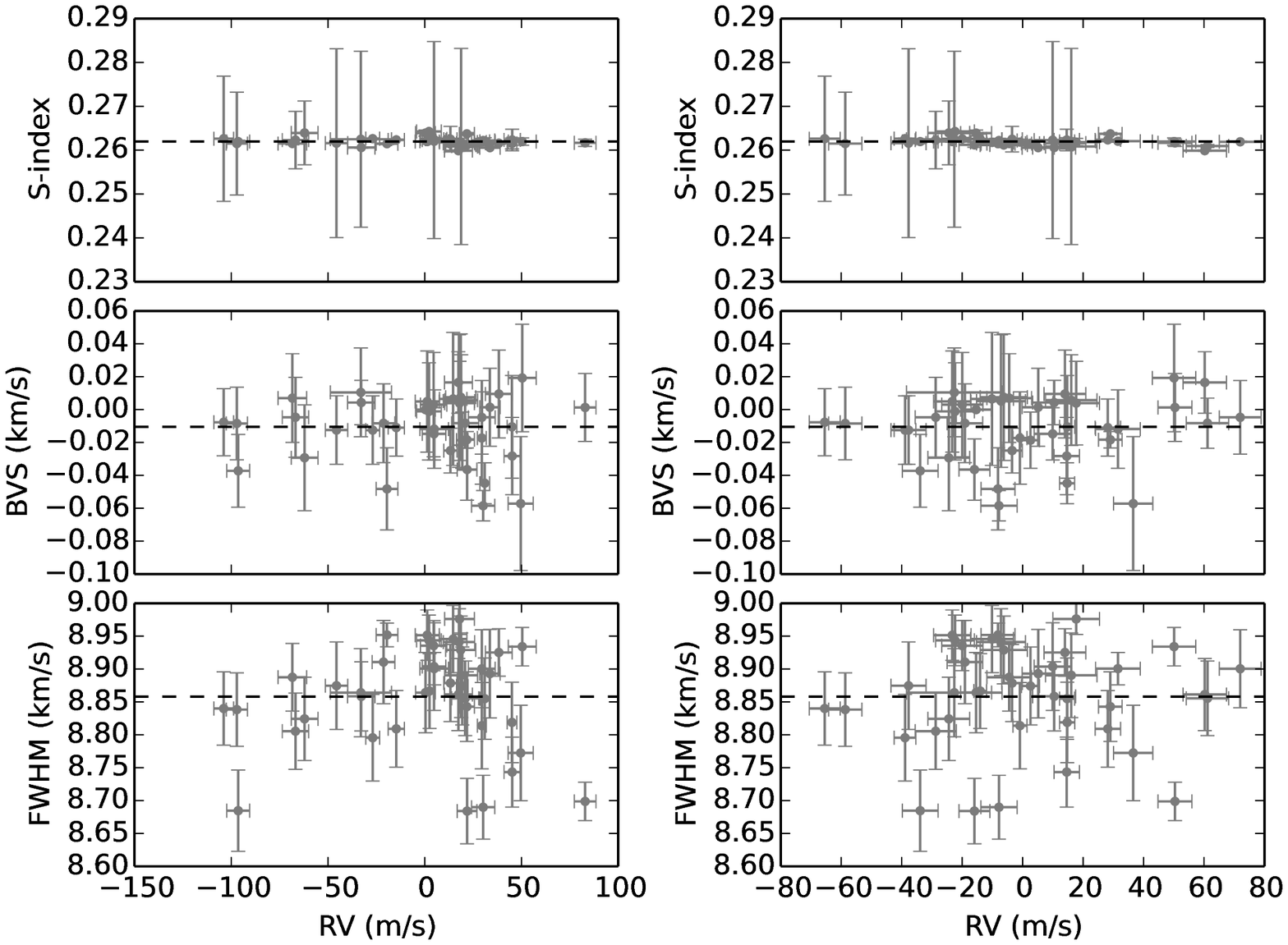}
\caption{S-index (upper panel), bisector velocity span (middle panel) and FWHM of the CCF (lower panel) versus the measured RV (left panel) and versus the residual RV for HD 122430, after fitting the best 1-planet solution we were able to obtain. Only the values for the activity indices laying 2.5$\sigma$ from the mean value of the distribution were considered. The correlation coefficients for the plots in the left panels are $r$= -0.17, 0.01 and 0.04, and for the right panels are $r=-0.32,$ 0.1 and 0.04 for the S-index, BVS and FWHM versus the residual velocity, respectively. Horizontal dashed lines correspond to the mean of the distribution for each plot.}
\label{HD122430_activity}
\end{figure}

\end{document}